\documentclass[reprint,prd,onecolumn,notitlepage,11pt]{revtex4-1}
\usepackage[english]{babel}
\usepackage{amsmath,amssymb,graphicx,bm,lipsum}
\usepackage[colorlinks=true,citecolor=blue,linkcolor=blue,urlcolor=blue]{hyperref} 
\usepackage[font={small},flushleft,indent]{caption}
\usepackage{setspace}
\usepackage{multirow}

\begin{document}

\title{Finite velocity effect of ultralight vector dark matter in pulsar timing arrays}

\author{Qing-Hua Zhu}
\email{zhuqh@cqu.edu.cn} 
\affiliation{Department of Physics and Chongqing Key Laboratory for Strongly Coupled Physics, Chongqing University, Chongqing 401331, China}
 
\begin{abstract} 

Pulsar timing array (PTA) collaborations have recently reported evidence of the stochastic gravitational waves.  In addition to gravitational wave signals from cosmological or astrophysical origins, PTAs are also capable of detecting coherently oscillating ultralight dark matter.   
This study investigates the PTA response to ultralight vector dark matter, incorporating the realistic finite velocity effects ($v\sim 10^{-3}$)  in our galaxy. 
Using a rigorous calculation framework, we demonstrate that both the timing residual amplitude (for deterministic signals) and the angular correlations (for stochastic backgrounds) depend sensitively on individual pulsar distances. Significantly, we find that the timing residuals are dominated by the Newtonian potential, contradicting prior studies which identified the curvature perturbation and tensor part of metric perturbation as the leading contributions. As expected, the PTA response to vector dark matter exhibits a characteristic anisotropic signature, making it distinguishable from the isotropic signal of scalar dark matter.

\end{abstract} 

\maketitle

\section{Introduction}

Pulsar timing array (PTA) collaborations have recently reported evidence of stochastic gravitational waves \cite{Xu:2023wog,Zic:2023gta,EPTA:2023akd,EPTA:2023fyk,Reardon:2023gzh,NANOGrav:2023gor,NANOGrav:2023tcn}. Because the pulse arrival time from pulsars can be influenced by spacetime fluctuations \cite{Detweiler:1979wn}, it could be spatially correlated \cite{Hellings:1983fr}. In the PTA frequency band, besides gravitational wave signals originating from cosmological sources \cite{EPTA:2023xxk,NANOGrav:2023hvm} or those generated by supermassive black hole binaries \cite{NANOGrav:2023hfp,NANOGrav:2023pdq,EPTA:2023xxk}, PTAs are also capable of detecting coherently oscillating ultralight dark matter in our galaxy \cite{Hu:2000ke, Khmelnitsky:2013lxt,Nomura:2019cvc,Wu:2023dnp,  EuropeanPulsarTimingArray:2023egv,EPTA:2023xxk,NANOGrav:2023hvm}.

Dark matter plays a crucial role in cosmology and astrophysical physics, as it provides interpretations of the gravitational effects that can not be accounted for by normal matter alone, such as those in rotation curves of galaxies \cite{Sofue:2000jx}, the formation of large-scale structures \cite{WMAP:2012nax}, and the cosmic microwave background \cite{Planck:2018vyg}. However, at the sub-galactic scale, there are observational discrepancies from the predictions of the standard cold dark matter model \cite{Weinberg:2013aya}. To address the small-scale controversies, the wave effect of dark matter was introduced, leading to the development of the fuzzy or ultralight dark matter \cite{Hu:2000ke,Ferreira:2020fam}. Recently, it has been associated with PTA observations \cite{Khmelnitsky:2013lxt,Porayko:2014rfa,Porayko:2018sfa,Kato:2019bqz,Kaplan:2022lmz}, suggesting that the dark matter scenario can be tested experimentally \cite{Xu:2023wog,Zic:2023gta,EPTA:2023fyk,NANOGrav:2023gor}.

The ultralight dark matter, composed of bosons with masses around $10^{-23}\text{eV}$, could behave as a wave-like source, affecting pulsar timing in nHz frequency band \cite{Khmelnitsky:2013lxt}. Assuming that the wave-like dark matter is deterministic \cite{Khmelnitsky:2013lxt,Nomura:2019cvc,Sun:2021yra,Unal:2022ooa,Wu:2023dnp}, it would propagate along a specific direction $\hat{k}$. As a result, the induced timing residual amplitude might depend on the relative locations between the source $\hat{k}$ and a pulsar on the sky. Theoretically, there are ultralight scalar, vector, and tensor dark matter. The timing residual amplitude induced by scalar dark matter does not have directional dependence \cite{Khmelnitsky:2013lxt}, whereas those induced by vector or tensor dark matter were shown to be directional dependent \cite{Nomura:2019cvc,Wu:2023dnp}. If the source has a stochastic nature \cite{Omiya:2023bio,Cai:2024thd,Armaleo:2020yml,Kim:2023kyy}, it should be analyzed statistically across the sky, and the signals from a pulsar pair could be spatially correlated. Recent studies have identified characteristic angular correlation signatures for the ultralight vector and tensor dark matter \cite{Omiya:2023bio,Cai:2024thd}. 
 
Ultralight dark matter have both  gravitational effect \cite{Khmelnitsky:2013lxt,Aoki:2016mtn,Porayko:2018sfa,Kato:2019bqz,Nomura:2019cvc,PPTA:2022eul,Sun:2021yra,Unal:2022ooa,Wu:2023dnp}, and interaction with normal matter \cite{Graham:2015ifn,Armaleo:2020yml,Kaplan:2022lmz,PPTA:2021uzb,Sun:2021yra,Unal:2022ooa}, both of which can affect pulsar timing. In this study, we focus on the pure gravitational effect of the ultralight vector dark matter.  A recent study has shown that there is a missing effect for ultralight scalar dark matter when considering the finite dark matter velocity $v(=k/m\approx10^{-3})$ \cite{Zhu:2024lht}. In this finite velocity setup, the timing residuals from deterministic source and the angular correlation from stochastic source both shown to be sensitive to the distances between the earth and pulsars. 
In this paper, we extensively study the pulsar timing influenced by ultralight vector dark matter, by considering the finite dark matter velocity. We consider both the PTA responses to deterministic and stochastic sources. The spacetime geometry sourced by the vector field inevitably results in an anisotropic spacetime. In the context of PTAs, this is expected to have signatures in the perturbative regime, leading to the modulation of the timing residuals.


The rest of the paper is organized as follows. In Section \ref{II}, we present the dynamical evolution of metric perturbations induced by ultralight vector dark matter, based on perturbing Einstein field equations to second order. In Section \ref{III}, we calculate the timing residuals and angular correlations originating from the ultralight vector dark matter and study their observational signatures. Finally, in Section \ref{IV}, conclusions and discussions are summarized.

\

\section{Spacetime fluctuations induced by oscillating ultralight dark matter\label{II}}
 
Ultralight dark matter, also known as fuzzy dark matter, was proposed to introduce wave effects on dark matter within our galaxy \cite{Hu:2000ke}. In this context, the wave-like dark matter can be assumed to freely propagate in the space, with pure gravitational effect. Thus, the Lagrange density of the vector dark matter takes the form of \cite{Nomura:2019cvc}
\begin{eqnarray}
  \mathcal{L} & = & - \frac{1}{4} F_{\mu \nu} F^{\nu \nu} - \frac{1}{2} m^2
  A_{\mu} A^{\mu} ~, \label{action}
\end{eqnarray}
where $F_{\mu \nu} = \partial_{\mu} A_{\nu} - \partial_{\nu} A_{\mu}$. In PTA observation, the pulsars in our galaxy are nearly situated in a flat spacetime. Thus, we consider the perturbed Minkowski metric in Newtonian gauge as follows \cite{Mukhanov:1990me},
\begin{eqnarray}
  \textrm{d} s^2 & = & - (1 + 2\phi^{(1)} + \phi^{(2)}) \textrm{d} t^2 + \left( \delta_{i  
  j} (1- 2\psi^{(1)} - \psi^{(2)}) + h_{i   j}^{(1)}+\frac{1}{2} h_{i   j}^{(2)} \right) \textrm{d} x^i \textrm{d}
  x^j \nonumber \\ && + (2V_i^{(1)}+V_i^{(2)}) \textrm{d} t \textrm{d} x^i ~, \label{met}
\end{eqnarray}
where $ V_i^{(n)}$ is the $n$th-order vector perturbation, $ h_{i   j}^{(n)}$ is the $n$th-order tensor
perturbations, the $\phi^{(n)}$ and $\psi^{(n)}$ are $n$th-order Newtonian potential perturbation and curvature perturbation, respectively. The perturbed metric is given to the second order, because the spacetime fluctuations induced by the dark matter are of nonlinear effect of gravity \cite{Khmelnitsky:2013lxt,Zhu:2022bwf}.

By making use of Eq.~(\ref{action}), we obtain the energy-momentum tensor of the vector field $A_\mu$, namely,
\begin{eqnarray}
  T_{\mu \nu} & = & \nabla_{\lambda} A_{\nu} \nabla^{\lambda} A_{\mu} +
  \nabla_{\mu} A^{\lambda} \nabla_{\nu} A_{\lambda} - \nabla^{\lambda} A_{\nu}
  \nabla_{\mu} A_{\lambda} - \nabla^{\lambda} A_{\mu} \nabla_{\nu} A_{\lambda}
  + m^2 A_{\mu} A_{\nu} \nonumber\\
  &  & + \frac{1}{2} g_{\mu \nu} (\nabla_{\lambda} A_{\sigma} \nabla^{\sigma}
  A^{\lambda}-\nabla_{\lambda} A_{\sigma}
  \nabla^{\lambda} A^{\sigma}  - m^2 A_{\lambda} A^{\lambda}) ~,\label{EMT}
\end{eqnarray}
where $\nabla_\mu$ is covariant derivative with respect to metric in Eq.~(\ref{met}) and $\nabla^\mu \equiv g^{\mu\nu}\nabla_\nu$. In flat spacetime background, the vector field can be expanded as $A^{\mu} = A^{(1), \mu} + 1 /
2 A^{(2), \mu} +\mathcal{O} (3)$. Perturbing the equation of energy-momentum tensor conservation to the second order, one can obtain, $(\nabla_{\mu} T^{\mu   \nu})^{(1)} \equiv 0$, and
\begin{subequations}
  \begin{eqnarray}
    \frac{1}{2} (\nabla_{\mu} T^{\mu   t})^{(2)} & = & - \left(
    \partial_t A^{\left( \text{v} \right)}_j + \partial_j \partial_t B^{\left(
    \text{s} \right)} - \partial_j A^{\left( \text{s}\right)} \right) \nonumber \\ &&  \times \Big(
    m^2 A^{\left( \text{v} \right), j} + \partial_0^2 A^{\left( \text{v}
    \right), j} - \partial_0 \partial^j A^{\left( \text{s}\right)} + m^2 
    \partial^j B^{\left( \text{s} \right)} + \partial^j \partial_0^2 B^{\left(
    \text{s} \right)} - \Delta A^{\left( \text{v} \right), j} \Big) = 0 ~, \\
    \frac{1}{2} (\nabla_{\mu} T^{\mu   a})^{(2)} & = &  \left( m^2
    A^{\left( \text{v} \right), j} + \partial_0^2 A^{\left( \text{v} \right), j}
    - \partial^j \partial_0 A^{\left( \text{s}\right)} + m^2 \partial^j
    B^{\left( \text{s} \right)} + \partial^j \partial_0^2 A^{\left( \text{s}
    \right)} - \Delta A^{\left( \text{v} \right), j} \right) \nonumber \\ &&  \times \left( \partial_j
    A^{\left( \text{v} \right), a} - \partial^a A^{\left( \text{v} \right)}_j
    \right) \nonumber\\
    &  & + \left( m^2 A^{\left( \text{s}\right)} - \Delta A^{\left( \text{s}
    \right)} + \Delta \partial_0 B^{\left( \text{s} \right)} \right) \left(
    \partial_0 A^{\left( \text{v} \right), a} - \partial^a A^{\left( \text{s}
    \right)} + \partial^a \partial_0 B^{\left( \text{s} \right)} \right)
    \nonumber\\
    &  & + m^2 \left( - A^{\left( \text{s}\right)} + \Delta A^{\left( \text{s}
    \right)} \right) \left( A^{\left( \text{v} \right), a} + \partial^a
    B^{\left( \text{s} \right)} \right) = 0 ~,
  \end{eqnarray}\label{DelT}
\end{subequations}
where $\partial^j \equiv
\delta^{j   l} \partial_l$, and we have used helicity decomposition for the $A^{(1)}_j$. Namely, the vector field is rewritten as
\begin{eqnarray}
  A^{(1)}_t \equiv A^{(\text{s})}, &&
  A^{(1)}_j  =  A^{\left( \text{v} \right)}_j + \partial_j
  B^{\left( \text{s} \right)}~,
\end{eqnarray}
where $A^{(\text{v})}_j$ and $B^{(\text{s})}$ are transverse part and longitude part of spatial vector field $A_j^{(1)}$, respectively. 
There is no $A^{(2)}_\mu$ in Eqs.~(\ref{DelT}) due to $A^{(0)} =
0$. Associating Eqs.~(\ref{DelT}) with Lorentz gauge,
\begin{eqnarray}
  \partial_{\mu} A^{\mu} = - \partial_0 A^{\left( \text{s}
\right)} + \Delta B^{\left( \text{s} \right)} = 0~, \label{LG}
\end{eqnarray}
one can obtain three independent components of $A^{(1), \mu}$ satisfying the wave equations as follows, 
\begin{subequations}
  \begin{eqnarray}  
    \partial_0^2 A^{\left( \text{s}\right)} + (m^2 - \Delta) A^{\left( \text{s}
    \right)} & = & 0 ~,\\
    \partial_0^2 A^{\left( \text{v} \right)}_j + (m^2 - \Delta) A^{\left(
    \text{v} \right)}_j & = & 0~.  
  \end{eqnarray} \label{EoM}
\end{subequations}
Since the temporal vector field $A_t^{(1)}$ can be determined by $B^{(\text{s})}$ based on Eq.~(\ref{LG}), we re-express it in terms of the subscript s for simplicity.
Notably, the massive vector field is not gauge invariant. We adopt the Lorentz gauge following previous studies \cite{Nomura:2019cvc,Omiya:2023bio}, and expect that the field could serve as a low-energy effective particle in a phenomenological framework.
The wave equations in Eqs.~(\ref{EoM}) suggest that the vector field could oscillate both in the space and time. Different components of the vector field  also evolve independently.   The solutions of Eqs.~(\ref{EoM}) can be obtained in the form of plane wave expansion, namely, 
\begin{eqnarray}
  A^{(I)} & = & \int \frac{\textrm{d}^3 k}{(2 \pi)^3} \left\{
  A_{\textbf{k}}^{(I)} e^{- i   \left( w_k t - \textbf{k} \cdot
  \textbf{x} \right)} + A_{-\textbf{k}}^{(I)} e^{i   \left( w_k t -
  \textbf{k} \cdot \textbf{x} \right)} \right\} ~, 
  \label{8}
\end{eqnarray}
where $w_k = \sqrt{k^2 + m^2}$ and we let $A^{(I)}$ represent $A^\text{(s)}$ or $ A^\text{(v)}_j$. From Eq.~(\ref{8}), the Fourier modes of $A^{(I)}_{\textbf{k}}$ can be given by
\begin{eqnarray}
  \tilde{A}_\textbf{k}^{(I)}
  & = & 2
  A_{\textbf{k}}^{(I)} \cos (w_k t) ~.  \label{solA}
\end{eqnarray}
The $A^{(\text{v})}_{\textbf{k},j}$ can be projected onto polarization vectors with respect to the wavenumber $\textbf{k}$, whereas the $A^{(\text{s})}_{\textbf{k}}$ need not. 
For a monochromatic vector field propagating in the direction of $\hat{k}$, the solution of wave equations [Eqs.~(\ref{EoM})] reduce to Eq.~(\ref{solA}) with a constant $\textbf{k}$.

In order to calculate metric perturbation induced by the oscillating vector fields, one should perturb Einstein field equations up to the second order. In the first order, the metric perturbations vanish, $\phi^{(1)}=\psi^{(1)}=V_i^{(1)}=h_{ij}^{(1)}=0$, due to $T_{\mu\nu}^{(1)}=0$. Thus, for illustration, we let $V_i \equiv V_i^{(2)}$, $h_{ij} \equiv h_{i   j}^{(2)}$, $\phi=\phi^{(2)}$ and $\psi\equiv\psi^{(2)}$.  In the second-order, Einstein filed equations  can be obtained as follows \cite{Maggiore:2018sht,Mukhanov:1990me},
\begin{subequations}
  \begin{eqnarray}
    \frac{1}{4} (\partial_0^2 h_{a   b} - \Delta h_{a   b}) & = & \kappa
    \Lambda_{a   b}^{c   d} T_{c   d}^{(2)} ~, \\
    - \frac{1}{4} \partial_0V_a & = & \kappa \Delta^{- 1} (\delta_a^c - \partial_a
    \Delta^{- 1} \partial^c) \partial^d T_{c   d}^{(2)} ~, \\
    \partial_0^2{\psi} & = & \kappa \partial^c \Delta^{- 1} \partial^d T_{c d}^{(2)} ~, \\
    \frac{1}{2} (\psi - \phi) & = & \frac{\kappa}{2} \Delta^{- 1} (3
    \partial^c \Delta^{- 1} \partial^d - \delta^{c   d}) T_{c   d}^{(2)}
    ~, 
  \end{eqnarray}\label{Einstein}
\end{subequations}
where the Latin alphabets denote the spatial indices,  we have employed the helicity decomposition for the spatial components of Einstein field equations, $\Lambda_{a   b}^{c d}[=(\delta^c_a -\partial^c\Delta^{-1}\partial_a)(\delta^d_b -\partial^d\Delta^{-1}\partial_b)-(\delta^{cd} -\partial^c\Delta^{-1}\partial^d)(\delta_{ab} -\partial_a\Delta^{-1}\partial_b)/2]$ is transverse-traceless operator, $\kappa[\equiv 8\pi G]$ is Einstein gravitational constant, and the energy-momentum tensor is given by
\begin{eqnarray}
  T^{(2)}_{b   c} & \equiv & T^{\left( \text{v,v} \right)}_{b  
  c} + T^{\left( \text{s,s} \right)}_{b   c} + T^{\left( \text{s,v}
  \right)}_{b   c}~, \label{11}
\end{eqnarray}
and
\begin{subequations}
  \begin{eqnarray}
    T^{\left( \text{v,v} \right)}_{b   c} & = & m^2 A^{\left( \text{v}
    \right)}_b A^{\left( \text{v} \right)}_c - \partial_0 A^{\left( \text{v}
    \right)}_b \partial_0 A^{\left( \text{v} \right)}_c + \partial^j A^{\left(
    \text{v} \right)}_b \partial_j A^{\left( \text{v} \right)}_c - \partial^j
    A^{\left( \text{v} \right)}_b \partial_c A^{\left( \text{v} \right)}_j \nonumber \\ && -
    \partial^j A^{\left( \text{v} \right)}_c \partial_b A^{\left( \text{v}
    \right)}_j + \partial_b A^{\left( \text{v} \right), j} \partial_c A^{\left(
    \text{v} \right)}_j \nonumber\\
    &  & + \frac{1}{2} \delta_{b   c} \left( - m^2 A^{\left( \text{v}
    \right)}_j A^{\left( \text{v} \right), j} + \partial_0 A^{\left( \text{v}
    \right)}_j \partial_0 A^{\left( \text{v} \right), j} + \partial_j A^{\left(
    \text{v} \right)}_l \partial^l A^{\left( \text{v} \right), j} - \partial_l
    A^{\left( \text{v} \right)}_j \partial^l A^{\left( \text{v} \right), j}
    \right) ~, \label{12a}\\
    T^{\left( \text{s,s} \right)}_{b   c} & = & - \partial_b A^{\left(
    \text{s}\right)} \partial_c A^{\left( \text{s}\right)} + \partial_b
    \partial_0 B^{\left( \text{s} \right)} \partial_c A^{\left( \text{s}
    \right)} + m^2 \partial_b B^{\left( \text{s} \right)} \partial_c B^{\left(
    \text{s} \right)} + \partial_b A^{\left( \text{s}\right)} \partial_c
    \partial_0 B^{\left( \text{s} \right)} - \partial_b \partial_0 B^{\left(
    \text{s} \right)} \partial_c \partial_0 B^{\left( \text{s} \right)}
    \nonumber\\
    &  & + \frac{1}{2} \delta_{b   c} \Big( m^2 \left( A^{\left(
    \text{s}\right)} \right)^2 + \partial_j A^{\left( \text{s}\right)}
    \partial^j A^{\left( \text{s}\right)} - 2 \partial_j \partial_0 B^{\left(
    \text{s} \right)} \partial^j B^{\left( \text{s} \right)} \nonumber \\ && - m^2 \partial_j
    B^{\left( \text{s} \right)} \partial^j B^{\left( \text{s} \right)}  +
    \partial_j \partial_0 B^{\left( \text{s} \right)} \partial^j \partial_0
    B^{\left( \text{s} \right)} \Big) ~, \label{12b}\\
    T^{\left( \text{s,v} \right)}_{b   c} & = & \partial_0 A^{\left(
    \text{v} \right)}_c \partial_b A^{\left( \text{s}\right)} + \partial_0
    A^{\left( \text{v} \right)}_b \partial_c A^{\left( \text{s}\right)} + m^2
    \left( A^{\left( \text{v} \right)}_b \partial_c B^{\left( \text{s} \right)}
    + A^{\left( \text{v} \right)}_c \partial_b B^{\left( \text{s} \right)}
    \right) \nonumber \\ && - \partial_0 A^{\left( \text{v}\right)}_b \partial_c \partial_0
    B^{\left( \text{s} \right)} - \partial_0 A^{\left( \text{v} \right)}_c
    \partial_b \partial_0 B^{\left( \text{s} \right)} \nonumber\\
    &  & + \delta_{b   c} \left( - m^2 A^{\left( \text{v} \right), j}
    \partial_j B^{\left( \text{s} \right)} + \partial_0 A^{\left( \text{v}
    \right), j} \left( \partial_b \partial_0 B^{\left( \text{s} \right)} -
    \partial_b A^{\left( \text{s}\right)} \right) \right) ~. \label{12c}
  \end{eqnarray} \label{Tsv}
\end{subequations}
Eqs.~(\ref{Einstein}) show that the scalar, vector, and tensor metric perturbations all can be induced by the vector field $A^{(1), \mu}$.

Given the explicit energy-momentum tensors in Eqs.~(\ref{11})--(\ref{Tsv}) and evolution of the vector field in Eqs.~(\ref{solA}), the motion of equations for the metric perturbations in Eqs.~(\ref{Einstein}) can be solved using Green's function approaches. To avoid obscuring the dominant physical behavior, we present only the leading-order expansion for $k/m \ll 1$.
Specifically, the metric perturbations sourced by $T_{bc}^{\rm (v,v)}$ in Eq.~(\ref{12a}) are given by
\begin{subequations}
\begin{align}
  h^{(\text{v,v})}_{ab, \textbf{k}} &= -4 \kappa \int \frac{\mathrm{d}^3 p}{(2\pi)^3} \left\{ \Lambda_{ab}{}^{jl} \cos(w t) \, A^{(\text{v})}_{\textbf{k}-\textbf{p}, j} A^{(\text{v})}_{\textbf{p}, l} \right\}, \label{eq:h_vv} \\
  V^{(\text{v,v})}_{d, \textbf{k}} &= 8 i \kappa \left( \frac{m}{k} \right) \int \frac{\mathrm{d}^3 p}{(2\pi)^3} \left\{ \frac{(k^2 \delta_d^j - k_d k^j) k^l}{k^3} \sin(w t) \, A^{(\text{v})}_{\textbf{k}-\textbf{p}, j} A^{(\text{v})}_{\textbf{p}, l} \right\}, \label{eq:V_vv} \\
  \psi^{(\text{v,v})}_{\textbf{k}} &= \kappa \int \frac{\mathrm{d}^3 p}{(2\pi)^3} \left\{ \left( \frac{1}{2} \delta^{jl} - \frac{k^j k^l}{k^2} \right) \cos(w t) \, A^{(\text{v})}_{\textbf{k}-\textbf{p}, j} A^{(\text{v})}_{\textbf{p}, l} \right\}, \label{eq:psi_vv} \\
  \phi^{(\text{v,v})}_{\textbf{k}} &= 4 \kappa \left( \frac{m}{k} \right)^2 \int \frac{\mathrm{d}^3 p}{(2\pi)^3} \left\{ \left( -\delta^{jl} + \frac{3 k^j k^l}{k^2} \right) \cos(w t) \, A^{(\text{v})}_{\textbf{k}-\textbf{p}, j} A^{(\text{v})}_{\textbf{p}, l} \right\} . \label{eq:phi_vv}
\end{align}
\end{subequations}
Similarly, the metric perturbations sourced by $T_{bc}^{\rm (v,s)}$ in Eq.~(\ref{12b}) are given by
\begin{subequations}
\begin{align}
  h^{(\text{s,s})}_{ab, \textbf{k}} &= \kappa \left( \frac{m}{k} \right)^2 \int \frac{\mathrm{d}^3 p}{(2\pi)^3} \left\{ \frac{4 k^2 \Lambda_{ab}{}^{cm} p_c p_m}{p^2 |\textbf{k}-\textbf{p}|^2} \cos(w t) \, A^{(\text{s})}_{\textbf{k}-\textbf{p}} A^{(\text{s})}_{\textbf{p}} \right\}, \label{eq:h_ss} \\
  V^{(\text{s,s})}_{d, \textbf{k}} &= 8 i \kappa \left( \frac{m}{k} \right)^3 \int \frac{\mathrm{d}^3 p}{(2\pi)^3} \left\{  \frac{(\textbf{k}\cdot\textbf{p}) (k_d \textbf{k}\cdot\textbf{p} - k^2 p_d)}{k p^2 |\textbf{k}-\textbf{p}|^2} \sin(w t) \, A^{(\text{s})}_{\textbf{k}-\textbf{p}} A^{(\text{s})}_{\textbf{p}} \right\}, \label{eq:V_ss} \\
  \psi^{(\text{s,s})}_{\textbf{k}} &= \kappa \left( \frac{m}{k} \right)^2 \int \frac{\mathrm{d}^3 p}{(2\pi)^3} \left\{ \frac{2(\textbf{k}\cdot\textbf{p})^2 -  k^2 (p^2 + \textbf{k}\cdot\textbf{p})}{2p^2 |\textbf{k}-\textbf{p}|^2} \cos(w t) \, A^{(\text{s})}_{\textbf{k}-\textbf{p}} A^{(\text{s})}_{\textbf{p}} \right\}, \label{eq:psi_ss} \\
  \phi^{(\text{s,s})}_{\textbf{k}} &= 4 \kappa \left( \frac{m}{k} \right)^4 \int \frac{\mathrm{d}^3 p}{(2\pi)^3} \left\{ \frac{-3 (\textbf{k}\cdot\textbf{p})^2 + k^2 (p^2 + 2 \textbf{k}\cdot\textbf{p})}{p^2 |\textbf{k}-\textbf{p}|^2} \cos(w t) \, A^{(\text{s})}_{\textbf{k}-\textbf{p}} A^{(\text{s})}_{\textbf{p}} \right\}.\label{eq:phi_ss}
\end{align}
\end{subequations}
And the metric perturbations sourced by $T_{bc}^{\rm (v,v)}$ in Eq.~(\ref{12c}) are given by
\begin{subequations}
\begin{align}
  h^{(\text{s,v})}_{ab, \textbf{k}} &= -8 i \kappa \left( \frac{m}{k} \right) \int \frac{\mathrm{d}^3 p}{(2\pi)^3} \left\{ \frac{k_j \Lambda_{ab}{}^{jc} p_c}{p^2} \sin(w t) \, A^{(\text{s})}_{\textbf{k}-\textbf{p}} A^{(\text{v})}_{\textbf{p}, j} \right\}, \label{eq:h_sv} \\
  V^{(\text{s,v})}_{d, \textbf{k}} &= 8 \kappa \left( \frac{m}{k} \right)^2 \int \frac{\mathrm{d}^3 p}{(2\pi)^3} \left\{ \frac{1}{p^2} \left( \delta_d^j \textbf{k}\cdot\textbf{p} + k^j \left( p_d - \frac{2 k_d \textbf{k}\cdot\textbf{p}}{k^2} \right) \right) \cos(w t) \, A^{(\text{s})}_{\textbf{k}-\textbf{p}} A^{(\text{v})}_{\textbf{p}, j} \right\}, \label{eq:V_sv} \\
  \psi^{(\text{s,v})}_{\textbf{k}} &= i \kappa \left( \frac{m}{k} \right) \int \frac{\mathrm{d}^3 p}{(2\pi)^3} \left\{ \frac{k^j (k^2 - 2 \textbf{k}\cdot\textbf{p})}{k p^2} \sin(w t) \, A^{(\text{s})}_{\textbf{k}-\textbf{p}} A^{(\text{v})}_{\textbf{p}, j} \right\}, \label{eq:psi_sv} \\
  \phi^{(\text{s,v})}_{\textbf{k}} &= -8 i \kappa \left( \frac{m}{k} \right)^3 \int \frac{\mathrm{d}^3 p}{(2\pi)^3} \left\{ \frac{k^j (k^2 - 3 \textbf{k}\cdot\textbf{p})}{k p^2} \sin(w t) \, A^{(\text{s})}_{\textbf{k}-\textbf{p}} A^{(\text{v})}_{\textbf{p}, j} \right\}. \label{eq:phi_sv}
\end{align} \label{delta_g}
\end{subequations}
The oscillation frequency $w$ appearing in Eqs.~(\ref{delta_g}) takes the form of
\begin{eqnarray}
  w & \equiv & 2 m + \frac{p^2 + \left| \textbf{k} - \textbf{p} \right|^2}{2
  m}~. \label{w}
\end{eqnarray}
The subleading expansions of the metric perturbations are presented in Appendix~\ref{appC}.
The non-oscillatory and lower-frequency modes, i.e. $w=0$ and $w=(k^2-2\textbf{k}\cdot\textbf{p})/2m$, have been neglected in Eqs.~(\ref{delta_g}), because these features might lie outside the sensitive frequency window of PTAs in the $\text{nHz}$ band.

The $V_{b,\textbf{k}}$ and $\phi_\textbf{k}$ are enhanced as $k/m \rightarrow0$, compared to the $h_{ab,\textbf{k}}$ and $\psi_\textbf{k}$. Specifically, we obtain the leading-order scaling relation, namely, 
\begin{eqnarray}
  (k/m)^2 \phi_\textbf{k} \sim (k/m) V_{b,\textbf{k}} \sim  \psi_\textbf{k} \sim h_{ab, \textbf{k}}~.
\end{eqnarray}
This relation is independent of which specific energy-momentum tensor ($T_{ab}^{\rm (v,v)}$, $T_{ab}^{\rm (s,s)}$ or $T_{ab}^{\rm (s,v)}$) is considered in Eqs.~(\ref{Einstein}) and (\ref{Tsv}). This enhancement originates from the inverse Laplacian on the right hand side of Eqs.~(\ref{Einstein}), which introduces $k^{-2}$ factors in Fourier space.
These dominant components $V_a$ and $\phi$ were neglected in previous studies \cite{Nomura:2019cvc,Omiya:2023bio}, since $V_{a,\textbf{k}}$ and $\phi_\textbf{k}$ might be ill-defined at exactly $k/m =0$. Our approach, based on the expansion at finite $v[\simeq{k/m}]$, provides a more robust formalism. How the distinction arises is further presented in Appendix~\ref{C}.

\section{Pulsar timing array observations \label{III}}

In the framework of Einstein's theory of gravity, gravity bends light in the spacetime. For the same reason, in the presence of gravitational fluctuations between the pulsar and the earth, the pulse arrival time can be either delayed or advanced. This effect can be formulated by the perturbed geodesic equation as follows,
\begin{eqnarray}
  \delta P^{\mu} \partial_{\mu} \bar{P}^{\nu} + \bar{P}^{\mu} \partial_{\mu}
  \delta P^{\nu} + \eta^{\nu \rho} \left( \partial_{\mu} \delta g_{\lambda
  \rho} - \frac{1}{2} \partial_{\rho} \delta g_{\mu \lambda} \right)
  \bar{P}^{\mu} \bar{P}^{\lambda} & = & 0 ~, 
\end{eqnarray}
where $P^\mu$ is the 4-velocity describing radio beams from a pulsar.
Using the perturbed metric in Eq.~(\ref{met}), the redshift, denoted by $z$, can be quantified by the temporal component of above equations, namely, \cite{Khmelnitsky:2013lxt,Maggiore:2018sht,Nomura:2019cvc}
\begin{eqnarray}
  z&\equiv& \frac{\delta \omega}{\omega}\bigg|^O_E = \frac{1}{2}\left(\frac{\delta P^0}{\bar{P}^0}-\frac{1}{2}\delta g_{00}\right)\bigg|^O_E \nonumber \\ & = & \frac{1}{2}\left(\psi(x_O)-\psi(x_E)\right) - \frac{1}{4} \hat{n}^i
  \hat{n}^j \big(h_{i   j}(x_O)-h_{ij}(x_E)\big)  \nonumber \\ &&  + \frac{1}{2}\int^t_{t-L} \textrm{d} \bar{t} \left\{ \hat{n}^l \partial_l
  \left( \phi \left( \bar{t}, \textbf{x} (\bar{t}) \right) + \psi \left( \bar{t}, \textbf{x} (\bar{t})
  \right) + \hat{n}^i V_i \left( \bar{t}, \textbf{x} (\bar{t}) \right) - \frac{1}{2}
  \hat{n}^i \hat{n}^j h_{i   j} \left( \bar{t}, \textbf{x} (\bar{t}) \right)
  \right) \right\}~, \label{z}
\end{eqnarray}
where $\hat{n}^i \equiv - \textrm{d} x^i / \textrm{d} t = - \bar{P}^i / \bar{P}^0$ and $\textbf{x}(\bar{t})=-\hat{n}(\bar{t}-t)$. The frequency of radio beam is $\omega \equiv u_\mu P^\mu$, where $u_\mu [=(-1+(1/2)\delta g_{00}, \bm 0)]$ is the comoving 4-velocity of the pulsar or the earth in the perturbed background in Eq.~(\ref{met}). The emission event from a pulsar is set to be $x_E = (t - L, L \hat{n}^i)$, where $L$ represents the distance from the earth to the pulsar, and the receiving event on the earth is $x_E = (t, 0)$. The timing residual, which is the accumulation of the $z$, can be given by  $R=\int^t_0 z\textrm{d}t$ \cite{Khmelnitsky:2013lxt,Maggiore:2018sht}.
From Eq.~(\ref{z}), it shows that the quantity $z$ originates from both the metric perturbations at the locations of the pulsars and the earth, as well as the accumulation of these metric perturbations along the line-of-sight. These effects are analogous to the Sachs-Wolfe and integrated Sachs-Wolfe effects in the context of cosmology \cite{Maggiore:2018sht}.

By making use of Eqs.~(\ref{delta_g}) and (\ref{z}), we  rewrite the $z$ in the
form of $z = z^{\left( \text{v,v} \right)} + z^{\left( \text{s,s} \right)} +
z^{\left( \text{s,v} \right)}$, and 
\begin{subequations}
  \begin{eqnarray}
    z^{\left( \text{v,v} \right)} & = & \int_{- \infty}^{\infty} \textrm{d} f \int
    \textrm{d} \Omega \left\{ \int \frac{k^2 \textrm{d} k \textrm{d}^3 p}{(2 \pi)^6} \delta
    \left( f - \frac{w}{2 \pi} \right) \kappa A^{\left( \text{v}
    \right)}_{c, \textbf{k} - \textbf{p}} A^{\left( \text{v} \right)}_{d,
    \textbf{p}} \mathcal{I}^{\left( \text{v,v} \right), c   d} \left(
    \textbf{k}, \textbf{p},\hat{n} \right) \mathcal{K}(f,k,\hat{n}) \right\}~,\\
    z^{\left( \text{s,s} \right)} & = & \int_{- \infty}^{\infty} \textrm{d} f \int
    \textrm{d} \Omega \left\{ \int \frac{k^2 \textrm{d} k \textrm{d}^3 p}{(2 \pi)^6} \delta
    \left( f - \frac{w}{2 \pi} \right) \kappa A^{\left( \text{s}
    \right)}_{\textbf{k} - \textbf{p}} A^{\left( \text{s} \right)}_{\textbf{p}}
    \mathcal{I}^{\left( \text{s,s} \right)} \left( \textbf{k}, \textbf{p},\hat{n}
    \right) \mathcal{K}(f,k,\hat{n})    \right\}~, \\
    z^{\left( \text{s,v} \right)} & = & \int_{- \infty}^{\infty} \textrm{d} f \int
    \textrm{d} \Omega \left\{ \int \frac{k^2 \textrm{d} k \textrm{d}^3 p}{(2 \pi)^6} \delta
    \left( f - \frac{w}{2 \pi} \right) \kappa A^{\left( \text{s}
    \right)}_{\textbf{k} - \textbf{p}} A^{\left( \text{s} \right)}_{d,
    \textbf{p}} \mathcal{I}^{\left( \text{s,v} \right),   d} \left(
    \textbf{k}, \textbf{p},\hat{n} \right) \mathcal{K}(f,k,\hat{n}) \right\}~,
  \end{eqnarray} \label{zgeneral}
\end{subequations}
where  $\mathcal{K}(f,k,\hat{n})\equiv e^{- 2 \pi i   f   t} ( 1
- e^{i   L \left( 2 \pi   f + \textbf{k} \cdot \hat{n}
\right)} )$. In the narrow frequency band ($2\pi f\simeq 2m$), the expansion in powers of $k/m$ in Eqs.~(\ref{delta_g}) reveals that the leading-order contribution to the redshift is
\begin{eqnarray}
  z &=& \underset{\text{Leading Order}}{\underbrace{\frac{1}{2}\hat{n}^i\int^t_{t-L} \textrm{d} \bar{t} \, \partial_j \phi \left( \bar{t}, \textbf{x} (\bar{t}) \right)}} \nonumber \\ && \underset{\text{Subleading Order}}{+\underbrace{ 
   \frac{1}{2}\left(\psi(x_O)-\psi(x_E)\right)- \frac{1}{4} \hat{n}^i
  \hat{n}^j \big(h_{i   j}(x_O)-h_{ij}(x_E)\big)+\frac{1}{2}\hat{n}^j \hat{n}^i\int^t_{t-L} \textrm{d} \bar{t} \, \partial_j V_i \left( \bar{t}, \textbf{x} (\bar{t}) \right)}} ~, 
   \nonumber \\ 
   &&+(\text{higher-order terms})
   \label{20}
\end{eqnarray}
This is the main result of our study and differs from pioneering works \cite{Nomura:2019cvc,Omiya:2023bio}. How the distinction arises is presented in Appendix~\ref{C}. Using Eqs.~(\ref{delta_g}) [along with the next-to-leading-order expansion in Eqs.~(\ref{solM})], the explicit expressions of $\mathcal{I}^{(I,J)}$ in Eqs.~(\ref{zgeneral}) take the form of
\begin{subequations}
  \begin{eqnarray}
    \mathcal{I}^{\left( \text{v,v} \right), jl} \left( \textbf{k},
    \textbf{p},\hat{n} \right) &=&\left( \frac{m}{k} \right)^2 \left(\frac{\textbf{k}\cdot\hat{n}}{2\pi f+\textbf{k}\cdot\hat{n}}\right) \left(- \delta^{lj} + \frac{3 k^{j} k^{l}}{k^2}\right) \nonumber\\
    &&+2 \left( \frac{\textbf{k}\cdot\hat{n}}{2\pi f+\textbf{k}\cdot\hat{n}}\right) \left( \frac{m}{k} \right) \left( \frac{1}{k} \right)^3 (k^2 \delta _{d}{}^{j} - k_{d} k^{j}) k^{l} \hat{n}^{d} 
    \nonumber \\ &&
    +\frac{1}{8} \left(-  \delta^{lj} +   \frac{2k^{j} k^{l}}{k^2}\right) -\frac{1}{2}\Lambda^{jl}_{a b}\hat{n}^a\hat{n}^b
    \nonumber \\ && +\mathcal{O}\left(k/m\right)~,
  \end{eqnarray}
\vspace{-1.cm}
  \begin{eqnarray}
    \mathcal{I}^{\left( \text{s,s} \right)} \left( \textbf{k}, \textbf{p},\hat{n}
    \right)&=& \left( \frac{m}{k} \right)^4 \left(\frac{\textbf{k}\cdot\hat{n}}{2\pi f+\textbf{k}\cdot\hat{n}}\right) \frac{-3 \left(\textbf{k}\cdot\textbf{p}\right)^2 + k^2 (p^2 + 2 \textbf{k}\cdot\textbf{p})}{p^2|\textbf{k}-\textbf{p}|^2} \nonumber\\
    &&+\frac{2}{k} \left( \frac{\textbf{k}\cdot\hat{n}}{2\pi f+\textbf{k}\cdot\hat{n}}\right) \left( \frac{m}{k} \right)^3 \frac{\hat{n}^{d} \textbf{k}\cdot\textbf{p} (k_{d} \textbf{k}\cdot\textbf{p} - k^2 p_{d}) }{p^2|\textbf{k}-\textbf{p}|^2} 
    \nonumber \\ &&
    +\frac{1}{8}\left(\frac{m}{k}\right)^2 \frac{k^2p^2+k^2 \textbf{k}\cdot\textbf{p}-2(\textbf{k}\cdot\textbf{p})^2+k^2\Lambda^{cd}_{ab}\hat{n}_c\hat{n}_d p^a p^b}{p^2|\textbf{k}-\textbf{p}|^2}
    \nonumber \\ && +\mathcal{O}\left({m}/{k}\right) ~,
  \end{eqnarray}
\vspace{-1.cm}
  \begin{eqnarray}
    \mathcal{I}^{\left( \text{s,v} \right),   j} \left( \textbf{k},
    \textbf{p},\hat{n} \right) &=& - \frac{2}{kp^2} \left( \frac{m}{k} \right)^3 \left(\frac{\textbf{k}\cdot\hat{n}}{2\pi f+\textbf{k}\cdot\hat{n}}\right)  k^{j} (k^2 -3 \textbf{k}\cdot\textbf{p}) \nonumber\\
    && +\frac{2}{k^2p^2} \left( \frac{\textbf{k}\cdot\hat{n}}{2\pi f+\textbf{k}\cdot\hat{n}}\right) \left( \frac{m}{k} \right)^2  \hat{n}^{d} (k^2 \delta _{d}{}^{j} \textbf{k}\cdot\textbf{p} + k^{j} (-2 k_{d} \textbf{k}\cdot\textbf{p} + k^2 p_{d}))    \nonumber \\ && 
    - \left(\frac{m}{k}\right) \frac{k^j (k^2-2\textbf{k}\cdot\textbf{p})}{4k p^2}- \left(\frac{m}{k} \right) \frac{k \Lambda_{c d}^{bj}\hat{n}^c\hat{n}^d p_b}{ p^2}
    \nonumber \\ &&+\mathcal{O}\left(1\right)~. \label{23c}
  \end{eqnarray} \label{22eq}
\end{subequations}
Because the observable in PTAs is the signal strain for a given frequency $f$, we have transformed the metric perturbation in Eq.~(\ref{delta_g}) into the frequency and direction $(f,\hat{k})$. We only consider the positive frequency parts of $z$ where $w>0$. 
In the subsequent sections, we will study the timing residuals from deterministic sources and the angular correlation of pulsar pairs from stochastic sources.

\

\subsection{Timing residuals \label{IIIA}}
 
For the oscillating vector dark matter propagating with a specific wavenumber
$\textbf{k}_{\ast}$ formulated by Eq.~(\ref{solA}), it can be understood as a type of deterministic source given by
\begin{subequations}
  \begin{eqnarray}
    A_{\textbf{k}, c}^{\left( \text{v} \right)} & = & \mathcal{A}^{\left(\text{v} \right)}_{\textbf{k}_\ast,\lambda} \epsilon_c^{\lambda} (\hat{k}) (2 \pi)^3 \delta^3
    \left( \textbf{k} - \textbf{k}_{\ast} \right)~, \\
    A_{\textbf{k}}^{\left( \text{s} \right)} & = & \mathcal{A}_{\textbf{k}_\ast}^{\left( \text{s}
    \right)} (2 \pi)^3 \delta^3 \left( \textbf{k} - \textbf{k}_{\ast} \right) ~,
  \end{eqnarray} \label{Dsrc}
\end{subequations}
where $\epsilon^\lambda$ is the polarization vector. It can serve as a set of orthogonal and normalized vectors spanning a three-dimensional vector space, namely,
  $\delta_{a   b}  =  \epsilon_a^1 \epsilon_b^1 + \epsilon_a^2
  \epsilon_b^2 + \hat{k}_a \hat{k}_b$.
Substituting Eqs.~(\ref{Dsrc}) into Eqs.~(\ref{zgeneral}), we obtain the real part of $z$ in the form of
\begin{subequations}
  \begin{eqnarray}
    z^{\left( \text{v,v} \right)} & = & \kappa \mathcal{A}^{\left( \text{v}
    \right)}_{{\textbf{k}_\ast},\lambda} \mathcal{A}^{\left( \text{v} \right)}_{{\textbf{k}_\ast},\lambda'}
    \epsilon_c^{\lambda} \epsilon_d^{\lambda'} \mathcal{I}^{\left( \text{v,v}
    \right), c   d} \left( 2 \textbf{k}_{\ast}, \textbf{k}_{\ast},\hat{n} \right)
    \left( \cos (2 \pi f_\ast  t) - \cos \left( 2 \pi f_\ast  t - L \left(
    2 \pi f_\ast+ \textbf{k} \cdot \hat{n} \right) \right) \right)~,\\
    z^{\left( \text{s,s} \right)} & = & \kappa \mathcal{A}^{\left( \text{s}
    \right)}_{\textbf{k}_\ast} \mathcal{A}^{\left( \text{s} \right)}_{\textbf{k}_\ast} \mathcal{I}^{\left(
    \text{s,s} \right)} \left( 2 \textbf{k}_{\ast}, \textbf{k}_{\ast},\hat{n} \right)
    \left( \cos (2 \pi f_\ast  t) - \cos \left( 2 \pi f_\ast  t - L \left(
    2 \pi f_\ast+ \textbf{k} \cdot \hat{n} \right) \right) \right)~,\\
    z^{\left( \text{s,v} \right)} & = & \kappa \mathcal{A}^{\left( \text{s}
    \right)}_{\textbf{k}_\ast} \mathcal{A}^{\left( \text{v} \right)}_{{\textbf{k}_\ast},\lambda}
    \epsilon_c^{\lambda} \mathcal{I}^{\left( \text{s,v} \right), c} \left( 2
    \textbf{k}_{\ast}, \textbf{k}_{\ast},\hat{n} \right) \left( \cos (2 \pi f_\ast 
    t) - \cos \left( 2 \pi f_\ast  t - L \left( 2 \pi f_\ast+ \textbf{k} \cdot
    \hat{n} \right) \right) \right)~,
  \end{eqnarray} \label{z1}
\end{subequations}
where $f_\ast\equiv \frac{1}{2 \pi} \left( 2 m + \frac{k_{\ast}^2}{m} \right)$, and
\begin{subequations}
  \begin{eqnarray} 
    \mathcal{I}^{\left( \text{v,v} \right), c   d} \left( 2
    \textbf{k}_{\ast}, \textbf{k}_{\ast},\hat{n} \right) & = & - \frac{1}{4} \left(
    \frac{m}{k_\ast} \right)^2 \left(\frac{\textbf{k}_{\ast} \cdot
    \hat{n}}{\pi f_\ast+\textbf{k}_{\ast} \cdot \hat{n} } \right) \delta^{c  
    d}  +\mathcal{O} \left( 1 \right)~,\\
    \mathcal{I}^{\left( \text{s,s} \right)} \left( 2 \textbf{k}_{\ast},
    \textbf{k}_{\ast},\hat{n} \right) & = & \frac{1}{2}\left( \frac{m}{k_\ast} \right)^4 \left(
    \frac{\textbf{k}_{\ast} \cdot \hat{n}}{\pi f_\ast+ \textbf{k}_{\ast} \cdot \hat{n}} \right)  +\mathcal{O} \left( \left( {m}/{k_\ast} \right)^2 \right)~,\\
    \mathcal{I}^{\left( \text{s,v} \right), d} \left( 2 \textbf{k}_{\ast},
    \textbf{k}_{\ast},\hat{n} \right) & = &  \left( \frac{m}{k_\ast} \right)^2 \left(
    \frac{\textbf{k}_{\ast} \cdot \hat{n}}{\pi f_\ast + \textbf{k}_{\ast} \cdot \hat{n}} \right) \hat{n}^d +\mathcal{O} \left(  {m}/{k_\ast} 
    \right)~. \label{26c}
  \end{eqnarray}\label{zI}
\end{subequations}
Given $\epsilon_c^\lambda k^c=0$ and the constraints specified in Eqs.~(\ref{Dsrc}), the first term of Eq.~(\ref{23c}) vanishes, reducing it to the form of Eq.~(\ref{26c}).
We consider unpolarized modes of $\mathcal{A}^{\left( \text{v}
\right)}_{\lambda}$, indicating
$\overline{\mathcal{A}_1^{\left( \text{v} \right)} \mathcal{A}_2^{\left(
\text{v} \right)}} = 0$ and $\overline{\mathcal{A}_1^{\left( \text{v}
\right)}} \simeq \overline{\mathcal{A}_2^{\left( \text{v} \right)}} \simeq \left| A^{(\text{v})}_{\bm k_\ast}\right|/\sqrt{2}$. Employing the spherical coordinate, $(\theta,\varphi) $, defined with
\begin{eqnarray}
  \epsilon^1 \cdot \hat{n}  =  \cos \varphi \sin \theta~, \hspace{0.5cm}
  \epsilon^2 \cdot \hat{n}  =  \sin \varphi \sin \theta~, \hspace{0.5cm} 
  \hat{k} \cdot \hat{n}  =  \cos \theta~,
\end{eqnarray}
we can evaluate Eq.~(\ref{z1}) to be
\begin{eqnarray}
  z & = & z^{\left( \text{v,v} \right)} + z^{\left( \text{s,s} \right)} +
  z^{\left( \text{s,v} \right)} = \kappa \mathcal{Z} \left( \cos (2 \pi f_\ast
    t) - \cos \left( 2 \pi f_\ast   t - L \left( 2 \pi f_\ast + \textbf{k}_\ast
  \cdot \hat{n} \right) \right) \right)~,
\end{eqnarray}
where
\begin{eqnarray} 
  \mathcal{Z} & \approx & \left(\frac{2    \cos \theta}{2 + v^2 + 2
  v   \cos \theta} \right) \left( - \frac{1}{2 v} \left| \mathcal{A}_{\textbf{k}_\ast}^{\left( \text{v} \right)}
  \right|^2  +  \frac{1}{v^3} \left( \mathcal{A}_{\textbf{k}_\ast}^{\left( \text{s} \right)}
  \right)^2 + \frac{\sqrt{2}}{v} \left| \mathcal{A}_{\textbf{k}_\ast}^{\left( \text{v} \right)} \right|
  \mathcal{A}_{\textbf{k}_\ast}^{\left( \text{s} \right)} \sin\theta (\cos\varphi + \sin\varphi) \right) ~.\label{29}
\end{eqnarray}  
Eq.~(\ref{29}) holds for a small dark matter speed $v \equiv k_{\ast} / m$.
The $\mathcal{Z}$ is a function of polarization angle $\varphi$ originating from the polarization vector of the vector field.  
Here, we have $\left| \mathcal{A}^{\left( \text{v} \right)}_{\textbf{k}_\ast}
\right| \equiv ( \delta^{\lambda \lambda'} \mathcal{A}_{{\textbf{k}_\ast},\lambda}^{\left(
\text{v} \right)} \mathcal{A}_{{\textbf{k}_\ast},\lambda'}^{\left( \text{v} \right)} )^{1
/ 2}$.  
And the $|\mathcal{A}_{\textbf{k}_\ast}^{(\text{v})}|$ and $(\mathcal{A}_{\textbf{k}_\ast}^{(\text{s})})$ are  set to be 
independent components. We have three possible cases: i) $ |\mathcal{A}_{\textbf{k}_\ast}^{(\text{v})}|\simeq \mathcal{A}_{\textbf{k}_\ast}^{(\text{s})}/v$, ii) $ |\mathcal{A}_{\textbf{k}_\ast}^{(\text{v})}|\gg \mathcal{A}_{\textbf{k}_\ast}^{(\text{s})}/v$, and iii) $ |\mathcal{A}_{\textbf{k}_\ast}^{(\text{v})}|\ll \mathcal{A}_{\textbf{k}_\ast}^{(\text{s})}/v$. In all these cases, we find that the $\mathcal{Z}$ is dominated by either the term $\mathcal{Z}^{(\text{v,v})}$ or $\mathcal{Z}^{(\text{s,s})}$. 

Following the approaches used in Ref.~\cite{Khmelnitsky:2013lxt}, the timing residuals of a pulsar can be computed by using $R = \int_0^t z 
\textrm{d} t$ and ignoring the non-oscillatory parts, namely,  
\begin{eqnarray}  
  R & = & \mathcal{R} \cos \left( 2\pi f_{\ast}   t - \frac{L}{2} \left(
  2\pi f_{\ast} + 2 \textbf{k}_{\ast} \cdot \hat{n} \right) \right) =\mathcal{R}
  \cos (2 \pi f_{\ast} t - \pi f_{\ast} L (1 + 2\alpha \cos \theta))~, 
\end{eqnarray}
where $\alpha \equiv k_{\ast} / (2 \pi f_{\ast})$, and the timing residual  amplitude is
\begin{eqnarray}
  \mathcal{R} & = &  \frac{\kappa\mathcal{Z}}{\pi f_\ast}   \sin (\pi f_{\ast} L
  (2\alpha \cos \theta - 1))~. \label{R}
\end{eqnarray}
In the limit $v \rightarrow 0$, the timing residual amplitude $\mathcal{R}$ is proportional to the direction-dependent factor $\mathcal{Z}$, which takes a dipole form. This is distinguishable from that of the scalar dark matter \cite{Khmelnitsky:2013lxt,Zhu:2024lht}.



\begin{figure}  
  \includegraphics[width=1\linewidth]{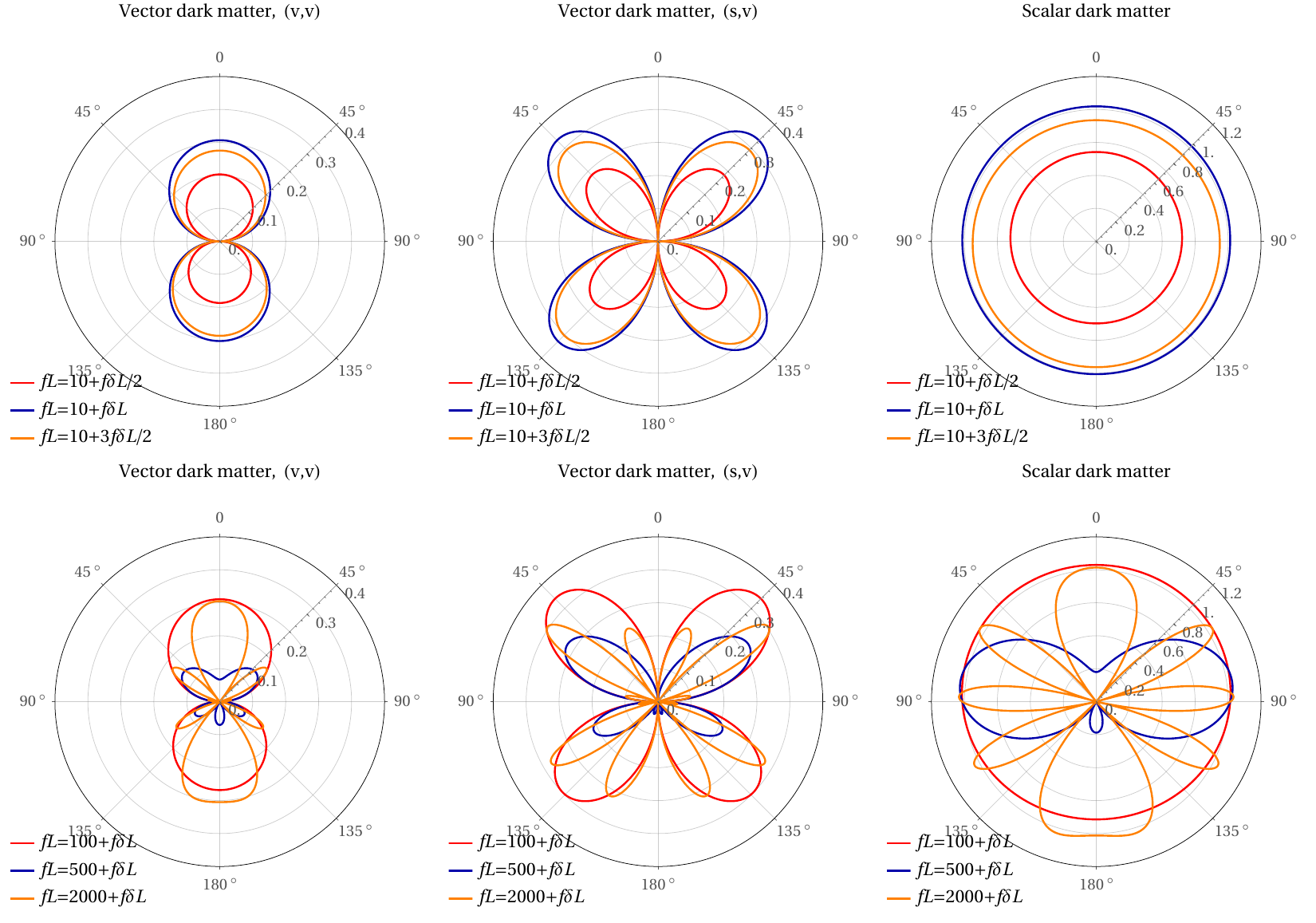}
  \caption{Comparison of direction-dependent timing residual amplitudes $\kappa^{-1} \pi f_\ast|\mathcal{R}|$ between scalar dark matter (see Ref.~\cite{Zhu:2024lht}) and vector dark matter for selected values of $fL$ in polar coordinates. Here, we set $v=0.001$, $\varphi=0$ and $|\mathcal{A}^{(\text{v})}_{\textbf{k}_\ast}|=\mathcal{A}^\text{(s)}_{\textbf{k}_\ast}=\mathcal{A}_\text{scalar}=1$. We did not plot the $\mathcal{R}^{{(\text{s,s})}}$, as $\mathcal{R}^{(\text{v,v})}$ and $\mathcal{R}^{(\text{s,s})}$ exhibit identical direction dependence at the leading order. \label{F1}}
\end{figure}

In Figure~\ref{F1}, we present a comparison of timing residual amplitudes for ultralight vector dark matter with those of scalar dark matter (see Ref.\cite{Zhu:2024lht}) with given pulsar distances. 
The dark matter speed is set to be $v\approx 0.001$. In the regime of $f_\ast L \alpha \ll 1$ as shown in the top panel of Figure~\ref{F1}, the factor $\sin (\pi f_{\ast} L  (2\alpha \cos \theta - 1))$ in Eq.~(\ref{R}) scales the timing-residual amplitude without altering the direction dependence. 
Here, the timing residual amplitude for scalar dark matter is isotropic, whereas that for vector dark matter exhibits characteristic direction dependence. Specifically, we have a dipole pattern for $\mathcal{R}^{(\text{v,v})}$ and $\mathcal{R}^{(\text{s,s})}$, and a quadrupole pattern for $\mathcal{R}^{(\text{s,v})}$.
In the regime $f_\ast L \alpha \gtrsim 1$, i.e., for realistic pulsar distances [$L\simeq (100$-$1000$)$f^{-1}$], the pattern of the timing residual amplitude exhibits an increased number of lobes, as shown in the bottom panel of Figure~\ref{F1}.

\

\subsection{Angular correlation}

Besides the ultralight vector dark matter $A^\mu$ being a potentially deterministic source formulated in Eq.~(\ref{Dsrc}), it also could have a stochastic nature \cite{Kim:2023kyy,Luu:2023rgg}. 
In this context, we set $A^{(\text{v})}_j$ and $A^{(\text{s})}$ both to be stationary, isotropic and Gaussian stochastic variables, and let $A^{(\text{v})}_j$ to be unpolarized. The statistical properties of the vector filed can be described by the two-point correlations, namely,
\begin{subequations}
  \begin{eqnarray}
    \langle A_{c, \textbf{k}'}^{\left( \text{v} \right)} A_{d,
    \textbf{k}}^{\left( \text{v} \right)} \rangle & = & \sum_{\lambda} (2
    \pi)^3 \delta^3 \left( \textbf{k}' + \textbf{k} \right) \epsilon_c^{\lambda}
    \epsilon_d^{\lambda} P^{\left( \text{v} \right)} (k) = (2 \pi)^3 \delta^3
    \left( \textbf{k}' + \textbf{k} \right) \mathcal{T}_{c   d} (\hat{k})
    P^{\left( \text{v} \right)} (k)~, \label{cor1}\\
    \langle A_{\textbf{k}'}^{\left( \text{s} \right)} A_{d,
    \textbf{k}}^{\left( \text{v} \right)} \rangle & = & 0~, \label{cor2}\\
    \langle A_{\textbf{k}'}^{\left( \text{s} \right)} A_{d,
    \textbf{k}}^{\left( \text{s} \right)} \rangle & = & (2 \pi)^3 \delta^3
    \left( \textbf{k}' + \textbf{k} \right) P^{\left( \text{s} \right)} (k)~,
  \end{eqnarray}\label{cor}
\end{subequations}
where $A^{\left( \text{v} \right)}_{j, \textbf{k}} = \epsilon_j^{\lambda}
A^{\lambda}_{j, \textbf{k}}$, and $\mathcal{T}_{c   d} \equiv \delta_{c
  d} - \hat{k}_c \hat{k}_d = \sum_{\lambda} \epsilon_c^{\lambda} \epsilon_d^{\lambda}$ is transverse operator. Eqs.~(\ref{cor1}) and (\ref{cor2}) indicate that the different components of the vector field are not correlated. 

By making use of Eqs.~(\ref{zgeneral}) and (\ref{cor}), the two-point correlation of $z$ can be obtained in the form of
\begin{eqnarray}
  \langle z_A z_{\textrm{B}} \rangle & = & \langle z_{\textrm{A}}^{\left(
  \text{v,v} \right)} z^{\left( \text{v,v} \right)}_{\textrm{B}} \rangle +
  \langle z_{\textrm{A}}^{\left( \text{s,s} \right)} z^{\left( \text{s,s}
  \right)}_{\textrm{B}} \rangle + \langle z_{\textrm{A}}^{\left( \text{s,v}
  \right)} z^{\left( \text{s,v} \right)}_{\textrm{B}} \rangle ~,
\end{eqnarray}
where
\begin{subequations}
  \begin{eqnarray}
    \langle z^{\left( \text{v,v} \right)}_{\textrm{A}} z^{\left( \text{v,v}
    \right)}_{\textrm{B}} \rangle & = & \kappa^2 \int_{- \infty}^{\infty}
    \textrm{d} f \int \textrm{d} \Omega \int \frac{k^2 \textrm{d} k \textrm{d}^3 p}{(2 \pi)^6}
    \Bigg\{ \delta \left( f - \frac{w}{2 \pi} \right)  
    \mathcal{T}_{c a} \left( \textbf{k} - \textbf{p} \right) 
    \mathcal{T}_{d   b} \left( \textbf{p} \right) \nonumber \\ && \times  \mathcal{I}^{\left(
    \text{v,v} \right), c   d} \left( \textbf{k}, \textbf{p},
    \hat{n}_{\textrm{A}} \right) \left( \mathcal{I}^{\left( \text{v,v} \right), a
      b} \left( \textbf{k}, \textbf{p}, \hat{n}_{\textrm{B}} \right)
    +\mathcal{I}^{\left( \text{v,v} \right), b   a} \left( \textbf{k},
    \textbf{k} - \textbf{p}, \hat{n}_{\textrm{B}} \right) \right) \nonumber \\ && \times   P^{\left( \text{v} \right)} \left( \left| \textbf{k} - \textbf{p}
    \right| \right) P^{\left( \text{v} \right)} (p) \left( 1 - e^{i  
    L_{\textrm{A}} \left( 2 \pi   f + \textbf{k} \cdot \hat{n}_{\textrm{A}}
    \right)} \right) \left( 1 - e^{- i   L_{\textrm{B}} \left( 2 \pi  
    f + \textbf{k} \cdot \hat{n}_{\textrm{B}} \right)} \right) \Bigg\}~,
  \end{eqnarray}  
  \vspace{-1.2cm}
  \begin{eqnarray}
    \langle z^{\left( \text{s,s} \right)}_{\textrm{A}} z^{\left( \text{s,s}
    \right)}_{\textrm{B}} \rangle & = & \kappa^2 \int_{- \infty}^{\infty}
    \textrm{d} f \int \textrm{d} \Omega \int \frac{k^2 \textrm{d} k \textrm{d}^3 p}{(2 \pi)^6}
    \Bigg\{ \delta \left( f - \frac{w}{2 \pi} \right) \nonumber \\ && \times  \mathcal{I}^{\left(
    \text{s,s} \right)} \left( \textbf{k}, \textbf{p}, \hat{n}_{\textrm{A}} \right)
    \left( \mathcal{I}^{\left( \text{s,s} \right)} \left( \textbf{k},
    \textbf{p}, \hat{n}_{\textrm{B}} \right) +\mathcal{I}^{\left( \text{s,s} \right)}
    \left( \textbf{k}, \textbf{k} - \textbf{p}, \hat{n}_{\textrm{B}} \right) \right)
    \nonumber \\ && \times  P^{\left( \text{s} \right)} \left( \left| \textbf{k} - \textbf{p}
    \right| \right) P^{\left( \text{s} \right)} (p) \left( 1 - e^{i  
    L_{\textrm{A}} \left( 2 \pi   f + \textbf{k} \cdot \hat{n}_{\textrm{A}}
    \right)} \right) \left( 1 - e^{- i   L_{\textrm{B}} \left( 2 \pi  
    f + \textbf{k} \cdot \hat{n}_{\textrm{B}} \right)} \right) \Bigg\} ~,
  \end{eqnarray}  
  \vspace{-1.2cm}
  \begin{eqnarray}
    \langle z^{\left( \text{s,v} \right)}_{\textrm{A}} z^{\left( \text{s,v}  \right)}_{\textrm{B}} \rangle & = & \kappa^2 \int_{- \infty}^{\infty}
    \textrm{d} f \int \textrm{d} \Omega \int \frac{k^2 \textrm{d} k \textrm{d}^3 p}{(2 \pi)^6}
    \Bigg\{ \delta \left( f - \frac{w}{2 \pi} \right)   \mathcal{T}_{c
      d}  \left( \textbf{p} \right) \nonumber \\ && \times \mathcal{I}^{\left( \text{s,v}
    \right),   d} \left( \textbf{k}, \textbf{p}, \hat{n}_{\textrm{A}} \right)
    \mathcal{I}^{\left( \text{s,v} \right), c} \left( \textbf{k}, \textbf{p},
    \hat{n}_{\textrm{B}} \right) 
    \nonumber \\ && \times  P^{\left( \text{s} \right)} \left( \left| \textbf{k} - \textbf{p}
    \right| \right) P^{\left( \text{v} \right)} (p) \left( 1 - e^{i  
    L_{\textrm{A}} \left( 2 \pi   f + \textbf{k} \cdot \hat{n}_{\textrm{A}}
    \right)} \right) \left( 1 - e^{- i   L_{\textrm{B}} \left( 2 \pi  
    f + \textbf{k} \cdot \hat{n}_{\textrm{B}} \right)} \right) \Bigg\}~.
  \end{eqnarray} \label{zz}
\end{subequations}
According to Eq.~(\ref{cor2}), the cross terms of the correlations should vanish, i.e., $\langle z^{\left( \text{v,v} \right)}_{\textrm{A}} z^{\left( \text{s,s}
  \right)}_{\textrm{B}} \rangle=\langle z^{\left( \text{v,v} \right)}_{\textrm{A}} z^{\left( \text{s,v}
  \right)}_{\textrm{B}} \rangle=\langle z^{\left( \text{s,s} \right)}_{\textrm{A}} z^{\left( \text{s,v}
  \right)}_{\textrm{B}} \rangle=0$.
Evaluating the volume element $\textrm{d}^3p$ as $\textrm{d}p\textrm{d}{|\textbf{k}-\textbf{p}|}\textrm{d}\varphi$, and adopting the dimensionless power spectrum $\mathcal{P}(k)=(k^3/(2\pi^2))P(k)$, Eqs.~(\ref{zz}) reduce to the form of
\begin{eqnarray}
  \langle z^{(I, J)}_{\textrm{A}} z^{(I, J)}_{\textrm{B}} \rangle & = &
  \frac{\kappa^2}{2} \int_{- \infty}^{\infty} \textrm{d} f \int \frac{\textrm{d}
  \Omega}{4 \pi} \int_0^{\infty} \textrm{d} k \int_0^{\infty} \textrm{d} p \int_{| k -
  p |}^{k + p} \textrm{d} \left| \textbf{k} - \textbf{p} \right| \Bigg\{
  \frac{k}{\left| \textbf{k} - \textbf{p} \right|^2 p^2} \delta \left( f -
  \frac{w}{2 \pi} \right) \nonumber \\ && \times  \mathcal{P}^{(I)} \left( \left| \textbf{k} -
  \textbf{p} \right| \right) \mathcal{P}^{(J)} (p)  (1 - e^{i   L_{\textrm{A}} (2 \pi   f + k (\hat{k} \cdot
  \hat{n}_{\textrm{A}}))}) (1 - e^{- i   L_{\textrm{B}} (2 \pi   f + k
  (\hat{k} \cdot \hat{n}_{\textrm{B}}))})\nonumber \\ && \times  \int \frac{\textrm{d} \varphi}{2 \pi}
  \mathcal{J}^{(I, J)} \left( \textbf{k}, \textbf{p}, \varphi ; \hat{n}_{\textrm{A}}, 
  \hat{n}_{\textrm{B}} \right) \Bigg\}~,
\end{eqnarray}
where indices $I$ and $J$ denote either s or v, and 
\begin{subequations}
  \begin{eqnarray}
    \int \frac{\textrm{d} \varphi}{2 \pi} \mathcal{J}^{\left( \text{v,v} \right)}
     & \equiv & \int \frac{\textrm{d} \varphi}{2 \pi} \Big\{ \mathcal{T}_{c a}
    \left( \textbf{k} - \textbf{p} \right) \mathcal{T}_{d   b} \left(
    \textbf{p} \right) \mathcal{I}^{\left( \text{v,v} \right), c   d}
    \left( \textbf{k}, \textbf{p}, \hat{n}_{\textrm{A}} \right) \nonumber \\ && \times \left(
    \mathcal{I}^{\left( \text{v,v} \right), a   b} \left( \textbf{k},
    \textbf{p}, \hat{n}_{\textrm{B}} \right) +\mathcal{I}^{\left( \text{v,v} \right),
    b   a} \left( \textbf{k}, \textbf{k} - \textbf{p}, \hat{n}_{\textrm{B}}
    \right) \right) \Big\}~,\\
    \int \frac{\textrm{d} \varphi}{2 \pi} \mathcal{J}^{\left( \text{s,s} \right)} & \equiv & \int \frac{\textrm{d} \varphi}{2 \pi} \left\{ \mathcal{I}^{\left(
    \text{s,s} \right)} \left( \textbf{k}, \textbf{p}, \hat{n}_{\textrm{A}}
    \right) \left( \mathcal{I}^{\left( \text{s,s} \right)} \left( \textbf{k},
    \textbf{p}, \hat{n}_{\textrm{B}} \right) +\mathcal{I}^{\left( \text{s,s} \right)}
    \left( \textbf{k}, \textbf{k} - \textbf{p}, \hat{n}_{\textrm{B}} \right)
    \right)\right\}~,\\
    \int \frac{\textrm{d} \varphi}{2 \pi} \mathcal{J}^{\left( \text{s,v} \right)}
     & \equiv & \int \frac{\textrm{d} \varphi}{2 \pi} \left\{ \mathcal{T}_{c  
    d} \left( \textbf{p} \right) \mathcal{I}^{\left( \text{s,v} \right),
      d} \left( \textbf{k}, \textbf{p}, \hat{n}_{\textrm{A}} \right)
    \mathcal{I}^{\left( \text{s,v} \right), c} \left( \textbf{k}, \textbf{p},
    \hat{n}_{\textrm{B}} \right) \right\}~.
  \end{eqnarray}\label{Intphi}
\end{subequations}
Eqs.~(\ref{Intphi}) can be formally calculated by making use of the formulae presented in Appendix~\ref{AppA}. 

The oscillating ultralight dark matter might be expected to be monochromatic, because metric perturbations are induced by its coherent oscillation \cite{Khmelnitsky:2013lxt}. In PTA observation, it was reported by NANOGrav collaborations that there is an indication of monopolar signal near 4 nHz frequency \cite{NANOGrav:2023hvm}. Therefore, we consider the monochromatic power spectrum in the form of
\begin{eqnarray}
  \mathcal{P}^{\left( \text{s} \right)} (k)  =  \mathcal{A}^{\left( \text{s}
  \right)} k_{\ast} \delta (k - k_{\ast})~, & &
  \mathcal{P}^{\left( \text{v} \right)} (k)  =  \mathcal{A}^{\left( \text{v}
  \right)} k_{\ast} \delta (k - k_{\ast})~. \label{35}
\end{eqnarray}
Using Eqs.~(\ref{35}), the two-point correlations in Eqs.~(\ref{zz}) can be expressed as
$\langle z^{(I, J)}_{\textrm{A}} z^{(I, J)}_{\textrm{B}} \rangle  = \int_{-
  \infty}^{\infty} \textrm{d} f \mathcal{S}(f) \Gamma(f, \bm x_\text{A}, \bm x_\text{B})$
where the power spectral density and the overlap reduction function take the following forms
\begin{eqnarray}
   \mathcal{S}^{(I, J)}(f) &=& \frac{\kappa^2}{2} \mathcal{A}^{(I)} \mathcal{A}^{(J)} \delta \left( f - \frac{1}{2 \pi} \left( 2 m +
  \frac{k_{\ast}^2}{m} \right) \right)~, \\
  \Gamma^{(I, J)}(f, \bm x_\text{A}, \bm x_\text{B}) &=&  \int
  \frac{\textrm{d} \Omega}{4 \pi} \int_{k_\text{Gal}/k_\ast}^2 \textrm{d} \tilde{k} \Bigg\{ \tilde{k} (1 -
  e^{2 \pi i   f   L_{\textrm{A}} (1+\alpha \tilde{k} (\hat{k}
  \cdot \hat{n}_{\textrm{A}}))}) (1 - e^{- 2 \pi i   f   L_{\textrm{B}}
  (1+\alpha \tilde{k} (\hat{k} \cdot \hat{n}_{\textrm{B}}))}) 
  \nonumber \\ && \times \left( \int \frac{\textrm{d} \varphi}{2 \pi} \mathcal{J}^{(I, J)} \left(
  \textbf{k}, \textbf{p}, \varphi ; \hat{n}_{\textrm{A}}, \hat{n}_{\textrm{B}} \right)
  \right)_{\left| \textbf{k} - \textbf{p} \right| = p = k_{\ast}} \Bigg\}_{2\pi f= k_\ast \alpha^{-1}}~. \label{zzdelta} 
\end{eqnarray}
Here, the scale of our galaxy $k_\text{Gal}$ serves an infrared cutoff following Ref.~\cite{Zhu:2024lht}, $\tilde{k} \equiv k / k_{\ast}$, and $\alpha \equiv k_{\ast} / 2 \pi f = ((2 m / k_{\ast}) + (k_{\ast} / m))^{- 1}$. The overlap reduction function $\Gamma(f, \bm x_\text{A}, \bm x_\text{B})$ depends on the locations of the pulsar pair, defined by $\bm x_\text{A}[=L_\text{A}\hat{n}_\text{A}]$ and $\bm x_\text{B}[=L_\text{B}\hat{n}_\text{B}]$, and characterizes the angular correlation of dark matter signals in PTAs.  Substituting Eqs.~(\ref{35}) into Eqs.~(\ref{Intphi}), we obtain the explicit expression,
\begin{subequations}
\begin{eqnarray}
  v^4\int \frac{\textrm{d} \varphi}{2 \pi} \mathcal{J}^{\left( \text{v,v} \right)} \Big|_{\left| \textbf{k} - \textbf{p} \right| = p = k_{\ast}}
  & = &  \frac{1}{2\tilde{k}^4} C_\text{1} (80 -16 \tilde{k}^2 + \tilde{k}^4)  
  - \frac{v^2}{4\tilde{k}^2} C_\text{2} (\tilde{k}^2 -12 ) 
   + \mathcal{O}(v^4)
  \\
  v^8\int \frac{\textrm{d} \varphi}{2 \pi} \mathcal{J}^{\left( \text{s,s} \right)}\Big|_{\left| \textbf{k} - \textbf{p} \right| = p = k_{\ast}} & = &  \frac{1}{2\tilde{k}^4} C_\text{1} (4 + \tilde{k}^2)^2 +
  \frac{v^2}{2\tilde{k}^2} C_\text{2} (4 + \tilde{k}^2)  
+ \mathcal{O}(v^4)  ~,\\
  v^6\int \frac{\textrm{d} \varphi}{2 \pi} \mathcal{J}^{\left( \text{s,v} \right)}\Big|_{\left| \textbf{k} - \textbf{p} \right| = p = k_{\ast}}
   & = &  - \frac{1}{\tilde{k}^2}C_\text{1} (-4 + \tilde{k}^2)  + \mathcal{O}(v^4) 
\end{eqnarray}
\end{subequations}
and 
\begin{eqnarray}
    C_\text{1}&\equiv& \left( \frac{\hat{k}\cdot\hat{n}_\text{A}}{\tilde{k}^{-1} \alpha^{-1}+\hat{k}\cdot\hat{n}_\text{A}} \right) \left( \frac{\hat{k}\cdot\hat{n}_\text{B}}{\tilde{k}^{-1} \alpha^{-1}+\hat{k}\cdot\hat{n}_\text{B}} \right)~.\\ 
  C_\text{2}&\equiv& \frac{\hat{k}\cdot\hat{n}_\text{A}}{\tilde{k}^{-1} \alpha^{-1}+\hat{k}\cdot\hat{n}_\text{A}}+\frac{\hat{k}\cdot\hat{n}_\text{B}}{\tilde{k}^{-1} \alpha^{-1}+\hat{k}\cdot\hat{n}_\text{B}}~, 
\end{eqnarray}
In the limit of $v\equiv k_\ast/m \rightarrow0$, the oscillating term in Eq.~(\ref{zzdelta}) can be factorized out due to $(e^{2 \pi i   f   L_{\textrm{A}} (1+\alpha \tilde{k} (\hat{k}\cdot \hat{n}_{\textrm{A}}))}) (1 - e^{- 2 \pi i   f   L_{\textrm{B}}(1+\alpha \tilde{k} (\hat{k} \cdot \hat{n}_{\textrm{B}}))}) \rightarrow (1 -e^{2 \pi i   f   L_{\textrm{A}}}) (1 - e^{- 2 \pi i   f   L_{\textrm{B}}})$.
However, in PTA observation, it expected that the ultralight vector dark matter oscillates in the PTA frequency band, $f\approx2m\simeq \mathcal{O}({1\text{nHz}})$, for distances $L\simeq \mathcal{O}(1\text{kpc})$, and with dark matter speed $v\simeq10^{-3}$. Hence, it leads to $2\pi fL\alpha \simeq \mathcal{O}(1)$, suggesting that the angular correlation can be affected by the distance of the pulsar to Earth, as shown in Sec.~\ref{IIIA}. 

\begin{figure} 
  \includegraphics[width=1.\linewidth]{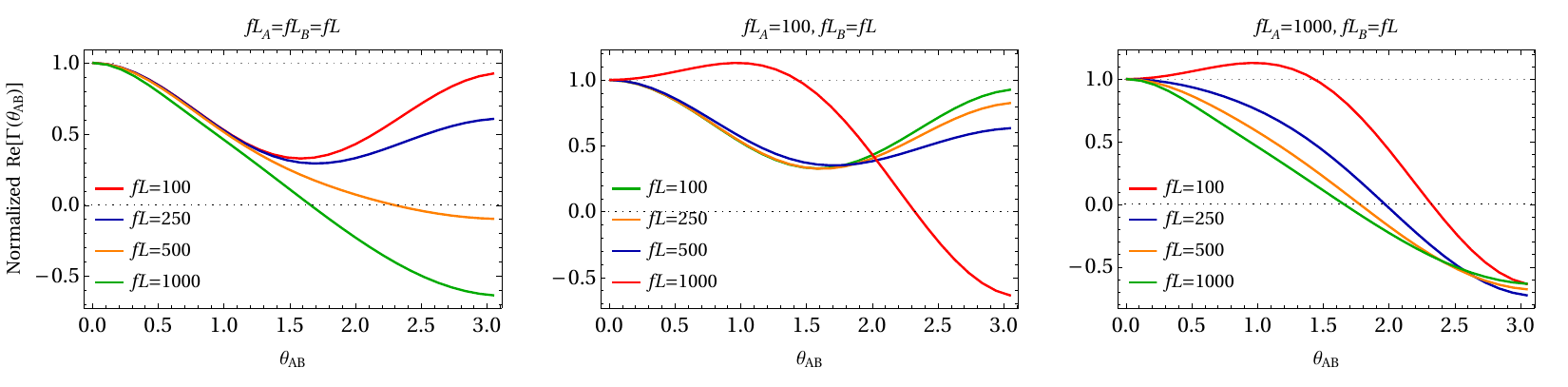} 
  \caption{Normalized angular correlations $\Gamma^{\text{(v,v)}}$ for selected $fL_\text{A}$ and $fL_\text{B}$. Here, we set $v=0.001$. \label{F2}}
\end{figure}

\begin{figure} 
  \includegraphics[width=1\linewidth]{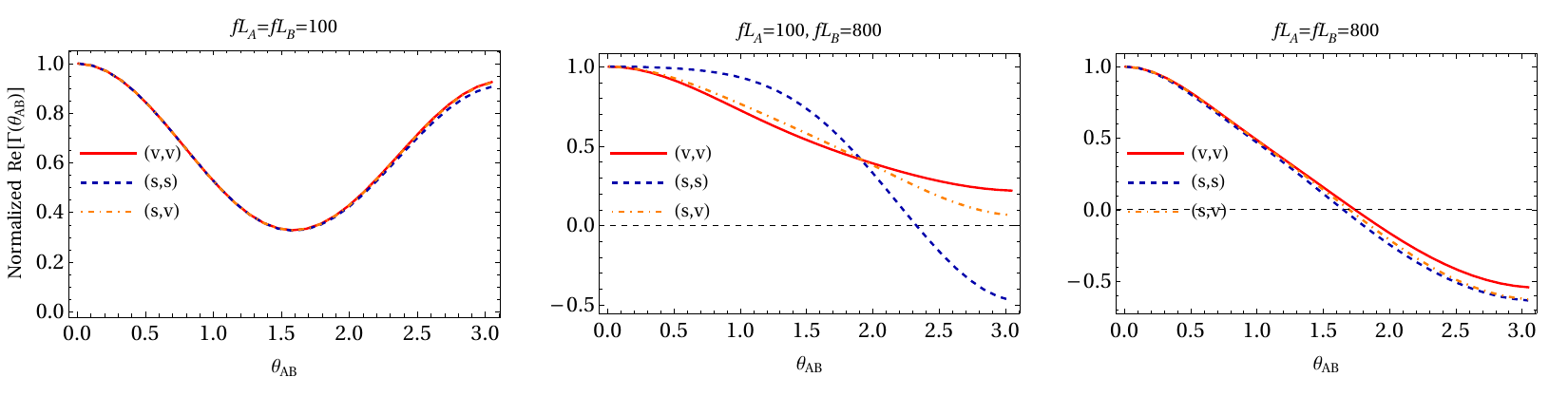}
  \caption{Normalized angular correlations $\Gamma^{\text{(v,v)}}$, $\Gamma^{\text{(s,s)}}$, and $\Gamma^{\text{(s,v)}}$  for given $fL$.  The $v$ is set to be $0.001$. 
   \label{F3}}
\end{figure}

\begin{figure}
  \includegraphics[width=1\linewidth]{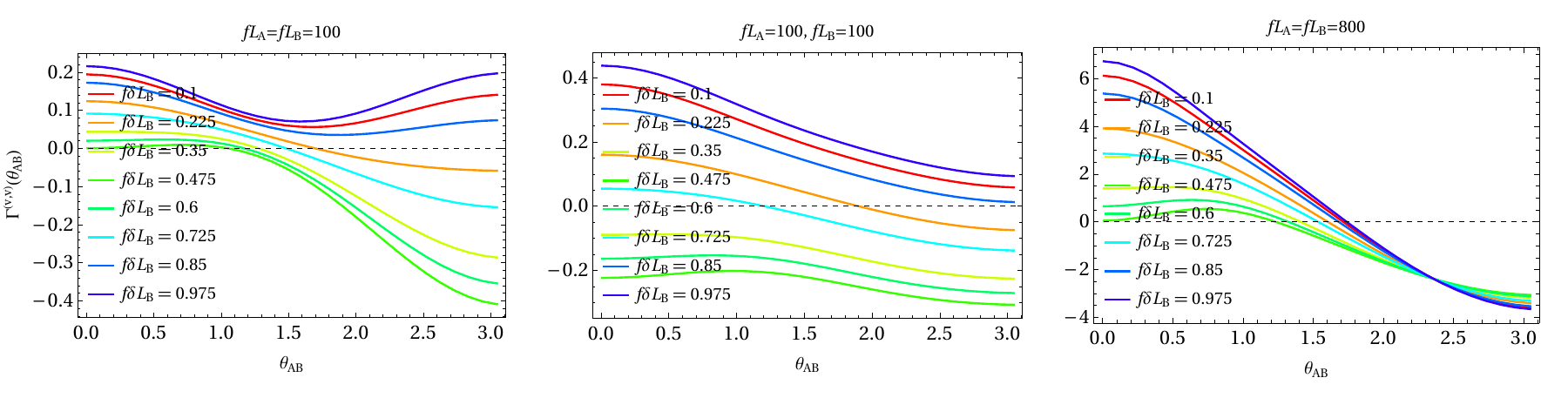}
  \caption{Normalized angular correlation $\Gamma^{\text{(v,v)}}$ with small variation of the distance $\delta L_\text{B}$ for given $fL_\text{A}$ and $fL_\text{B}$. \label{F4}}  
\end{figure}

In Figure~\ref{F2}, we present the angular correlation $\Gamma^\text{(v,v)}$ for selected values of $f L_\text{A}$ and $f L_\text{B}$, revealing that pulsar distances significantly affect the angular correlation. Figure~\ref{F3} compares the three mode combinations $\Gamma^\text{(v,v)}$, $\Gamma^\text{(s,s)}$, and $\Gamma^\text{(s,v)}$ for given $f L$. In the symmetric case $f L_\text{A}=f L_\text{B}$, these components exhibit similar angular profiles. However, different mode combinations become distinguishable when the pulsar pair has asymmetric distances ($f L_\text{A}\neq f L_\text{B}$). Figure~\ref{F4} further demonstrates the angular correlation's sensitivity to small distance variations $\delta L$, highlighting the importance of precise distance measurements.

\ 

\section{Discussions and conclusions \label{IV}}

This paper investigated the PTA response to the ultralight vector dark matter in our galaxy, incorporating the finite dark matter velocity in PTAs.
We revealed that both the timing residual amplitude from a deterministic source and the angular correlation from stochastic sources depend sensitively on individual pulsar distances. Based on a rigorous calculation framework, our results showed that timing residuals induced by vector dark matter are dominated by the Newtonian potential $\phi$ (Eq.~(\ref{20})). This finding contradicts previous studies, where the curvature perturbation $\psi$ and tensor perturbation $h_{ij}$  dominate the PTA response \cite{Nomura:2019cvc,Omiya:2023bio}. Notably, the PTA response function of vector dark matter exhibits a characteristic anisotropic signature, making it distinguishable from the typically isotropic signal of scalar dark matter \cite{Khmelnitsky:2013lxt,Zhu:2024lht}.

The timing residual amplitude in Eq.~(\ref{R}) is enhanced due to small dark matter speed $v\approx 10^{-3}$. It improves the observational feasibility of the vector dark matter, suggesting that tighter constraints might be placed on vector dark matter compared to scalar dark matter.  However, it might lead to concerns that the dark matter density is overestimated as the $v\rightarrow0$. We calculated dark matter density in Appendix~\ref{appB}, and demonstrated no such issues. In physical viewpoint, the vector dark matter speed $v$ should also be non-zero. Because the oscillating dark matter in PTA observations is confined to our galaxy, its wavelength can not exceed the galactic scale $\lambda_\text{galaxy}$. Specifically, there must be a natural infrared cutoff for the vector fields, such that $A_{j,\textbf{k}}=0$ for $k\lesssim2\pi\lambda_\text{galaxy}^{-1}$. In this sense, our calculations remain physically valid in the context of PTA observations.

We used helicity decomposition to evaluate the Einstein field equations and introduced different types of metric perturbations (also see Refs.~\cite{Zhu:2022wbd, Zhu:2022bwf, Zhu:2024lht}). This approach differs from previous studies that adopted the local frame \cite{Nomura:2019cvc, Omiya:2023bio, Wu:2023dnp}. As noted by Ashtekar and Bonga, the both approaches are used to introduce transverse-traceless modes, but ``conceptually distinct and the difference persists even in the asymptotic region'' \cite{Ashtekar:2017wgq}. Therefore, it expected that distinct approaches might lead to different physical effects also in PTA observations, and motivated to be addressed in future studies.

\ 

{\it Acknowledgments.} This work is supported by the National Natural Science Foundation of China under grants No.~12305073 and No.~12347101. We thank the anonymous referee for the valuable comments.

\

\appendix

\section{Azimuthal angle integration of multiple momentum $\textbf{p}$\label{AppA}}

In Cartesian coordinate, the momentum can be expressed in the form of
\begin{eqnarray}
  p^i & = & p (\sin \varphi \sin \theta \hat{x} + \cos \varphi \sin \theta \hat{y} +
  \cos \theta \hat{z})~.
\end{eqnarray}
By making use of above $p^i$, we obtain 
\begin{subequations}
  \begin{eqnarray}
    \int \frac{\textrm{d} \varphi}{2 \pi} p^i & = & p \cos \theta \hat{z}^i ~,\\
    \int \frac{\textrm{d} \varphi}{2 \pi} p^i p^j & = & \frac{1}{2} p^2 \sin^2 \theta
    \delta^{i   j} + \frac{1}{2} p^2 (3 \cos^2 \theta - 1) \hat{z}^i
    \hat{z}^j  \\
    \int \frac{\textrm{d} \varphi}{2 \pi} p^i p^j p^l & = & \frac{1}{2} p^3 \cos \theta
    \sin^2 \theta (\hat{z}^i \delta^{j   l} + \hat{z}^j \delta^{l i} +
    \hat{z}^l \delta^{i   j}) + \frac{1}{2} p^3 \cos \theta (5 \cos^2
    \theta - 3) \hat{z}^i \hat{z}^j \hat{z}^l~,  \\
    \int \frac{\textrm{d} \varphi}{2 \pi} p^i p^j p^l p^m & = & \frac{1}{8} p^4 \sin^4 
    \theta (\delta^{i   j} \delta^{l   m} + \delta^{i l} \delta^{j
      m} + \delta^{i   m} \delta^{l     j}) +
    \frac{1}{8} p^4 (35 \cos^4 \theta - 30 \cos^2 \theta + 3) \hat{z}^i
    \hat{z}^j \hat{z}^l \hat{z}^m \nonumber \\
    &  & + \frac{1}{8} p^4 \sin^2 \theta (5 \cos^2 \theta - 1) (\hat{z}^i
    \hat{z}^j \delta^{l   m} + \hat{z}^i \hat{z}^l \delta^{j   m}
    + \hat{z}^i \hat{z}^m \delta^{j   l} \nonumber\\ && + \hat{z}^j \hat{z}^l \delta^{i
      m} + \hat{z}^j \hat{z}^m \delta^{i   l} + \hat{z}^l
    \hat{z}^m \delta^{i   j})~.  
  \end{eqnarray} \label{A1}
\end{subequations}
By making use of Eqs~(\ref{A1}), we can calculate the integration over $\varphi$ in Eqs.~(\ref{Intphi}) with $\hat{z}\equiv\hat{k}$ and $\theta\equiv \arccos(\hat{k}\cdot\textbf{p}/p)=\arccos\left( (k^2+p^2-|\textbf{k}-\textbf{p}|^2)/(2kp) \right)$.

\section{Next-to-leading-order expansion of Eqs.~(\ref{delta_g}) \label{appC}}
Following the approaches used in Eqs.~(\ref{delta_g}), we expand the metric perturbations to next-to-leading order for small $v$, which can be given by
\begin{eqnarray}
  {\delta g}^{\left( {I,J} \right)}_{\mu\nu, \textbf{k}} & = &  \kappa
   \int \frac{\textrm{d}^3 p}{(2 \pi)^3}
  \left\{ \mathcal{M}^{\left( {I,J} \right)}_{\delta g,\mu\nu}(\textbf{k},\textbf{p};t,m) \textrm{trig}^{(I,J)}_{\delta g} (w t) A^{\left( I \right)}_{ \textbf{k} -
  \textbf{p}} A^{\left( J \right)}_{ \textbf{p}} \right\}~,\label{PsiIJ}
\end{eqnarray} 
where $\text{trig}^{(I,J)}_{\delta g}$ denotes a trigonometric function $\sin$ or $\cos$, the $I$ and $J$ represent the superscript v/s for the scalar/vector components of the vector fields, the quantity ${\delta g_{\mu\nu}}$ is the metric perturbations, and $w$ is defined in Eq.~(\ref{w}).
 For example, in the case of $I=J=\text{v}$ and ${\delta g}_{ab}=h_{ab}$, we have $\mathcal{M}^{\left( \text{v,v} \right),cd}_{h,ab}(\textbf{k},\textbf{p};t,m)\cos(w t) A^{\left( \text{v} \right)}_{ \textbf{k} -
  \textbf{p},c} A^{\left( \text{v} \right)}_{ \textbf{p},d}$ in Eq.~(\ref{PsiIJ}).  The explicit expressions of $\mathcal{M}_{\delta g}^{(I,J)}$ are presented as follows,
\begin{subequations}
  \begin{eqnarray}
    \mathcal{M}^{\left( \text{v,v} \right),jl}_{h,ab}(\textbf{k},\textbf{p};t,m) & = & -4 \Lambda_{ab}{}^{lj} - \left( \frac{k}{m} \right)^2 \frac{2  p_{c} (2 k^{l} \Lambda_{ab}{}^{cj} + \delta^{jl} \Lambda_{ab}{}^{cm} p_{m})}{k^2}+\mathcal{O}\left((k/m)^3\right)~, \label{15a}\\
    \mathcal{M}^{\left( \text{v,v} \right),jl}_{V,d}(\textbf{k},\textbf{p};t,m) & = & \left( \frac{m}{k} \right) \frac{8i  (k^2 \delta _{d}{}^{j} - k_{d} k^{j}) k^{l}}{k^3} + 4i \left( \frac{k}{m} \right) \left( \frac{1}{k} \right)^5 (k^2 \delta _{d}{}^{l} k^{j} (p^2 - \textbf{k}\cdot\textbf{p}) \nonumber \\ && - k_{d} (\left(\textbf{k}\cdot\textbf{p}\right)^2 \delta^{jl}  + k^{j} k^{l} (p^2 + \textbf{k}\cdot\textbf{p})) \nonumber \\ && + k^2 (\delta _{d}{}^{j} k^{l} \textbf{k}\cdot\textbf{p} + (k^{j} k^{l} + \delta^{jl} \textbf{k}\cdot\textbf{p}) p_{d}))+\mathcal{O}\left((k/m)^2\right) ~,\\
    \mathcal{M}^{\left( \text{v,v} \right),jl}_{\psi}(\textbf{k},\textbf{p};t,m) &=& \frac{1}{2} \delta^{jl} - \frac{k^{j} k^{l}}{k^2} + \frac{1}{8} \left( \frac{k}{m} \right)^2 \left( \frac{1}{k} \right)^4 (4 k^{j} k^{l} - \delta^{jl} (k^2 -2 \textbf{k}\cdot\textbf{p})) (k^2 -2 \textbf{k}\cdot\textbf{p}) \nonumber \\
    && +\mathcal{O}\left((k/m)^3\right) ~, \label{15c}\\
    \mathcal{M}^{\left( \text{v,v} \right),jl}_{\phi}(\textbf{k},\textbf{p};t,m) &=&4\left( \frac{m}{k} \right)^2 \left(- \delta^{jl} + \frac{3 k^{j} k^{l}}{k^2}\right) - \frac{1}{2} \left( \frac{1}{k} \right)^4 ((k^4 -12 \left(\textbf{k}\cdot\textbf{p}\right)^2 + 12 k^2 p^2) \delta^{jl} \nonumber \\ && + 4 (k^2 -6 p^2) k^{j} k^{l}) + \frac{1}{8} \left( \frac{k}{m} \right)^2 \left( \frac{1}{k} \right)^6 (k^2 -2 \textbf{k}\cdot\textbf{p}) (-2 k^{j} k^{l} (k^2 -6 \textbf{k}\cdot\textbf{p}) \nonumber \\ && + \delta^{jl} (k^4 -2 k^2 \textbf{k}\cdot\textbf{p}))+\mathcal{O}\left((k/m)^3\right)~, \label{15d}
  \end{eqnarray}
\vspace{-1.2cm}
  \begin{eqnarray}
    \mathcal{M}^{\left( \text{s,s} \right)}_{h,ab}(\textbf{k},\textbf{p};t,m) &=&\frac{1}{p^2|\textbf{k}-\textbf{p}|^2} \left(4 k^2 \left( \frac{m}{k} \right)^2 \Lambda_{ab}{}^{cm} p_{c} p_{m} + 2 \Lambda_{ab}{}^{cm} (- p^2 + \textbf{k}\cdot\textbf{p}) p_{c} p_{m}\right)\nonumber \\ && +\mathcal{O}\left(k/m\right)~,\\
    \mathcal{M}^{\left( \text{s,s} \right)}_{V,d}(\textbf{k},\textbf{p};t,m)&=& \frac{8i}{k} \left( \frac{m}{k} \right)^3 \frac{\textbf{k}\cdot\textbf{p} (k_{d} \textbf{k}\cdot\textbf{p} - k^2 p_{d})}{p^2|\textbf{k}-\textbf{p}|^2}+\mathcal{O}\left((k/m)^{-1}\right)  ~,\\
    \mathcal{M}^{\left( \text{s,s} \right)}_{\psi}(\textbf{k},\textbf{p};t,m)&=& \frac{1}{p^2|\textbf{k}-\textbf{p}|^2} \Big(\left( \frac{m}{k} \right)^2 (\left(\textbf{k}\cdot\textbf{p}\right)^2 - \frac{1}{2} k^2 (p^2 + \textbf{k}\cdot\textbf{p})) \nonumber\\ &&- \frac{(p^2 - \textbf{k}\cdot\textbf{p}) (3 k^4 + 4 \left(\textbf{k}\cdot\textbf{p}\right)^2 -8 k^2 \textbf{k}\cdot\textbf{p})}{8 k^2}\Big)+\mathcal{O}\left(k/m\right)~, \label{15g}\\
    \mathcal{M}^{\left( \text{s,s} \right)}_{\phi}(\textbf{k},\textbf{p};t,m) &=&\frac{1}{p^2|\textbf{k}-\textbf{p}|^2} \bigg(4 \left( \frac{m}{k} \right)^4 (-3 \left(\textbf{k}\cdot\textbf{p}\right)^2 + k^2 (p^2 + 2 \textbf{k}\cdot\textbf{p})) \nonumber\\ && - \frac{1}{8} \left( \frac{1}{k} \right)^4 (k^2 -2 \textbf{k}\cdot\textbf{p}) (12 \textbf{k}\cdot\textbf{p}^3 + k^4 (5 p^2 + \textbf{k}\cdot\textbf{p}) -6 k^2 \textbf{k}\cdot\textbf{p} (p^2 + 2 \textbf{k}\cdot\textbf{p})) \nonumber\\ && + \left( \frac{m}{k} \right)^2 \Big(\frac{6 \left(\textbf{k}\cdot\textbf{p}\right)^2 (- p^2 + \textbf{k}\cdot\textbf{p})}{k^2} + \frac{1}{2} k^2 (p^2 + 3 \textbf{k}\cdot\textbf{p}) \nonumber\\ && + 2 (-3 \left(\textbf{k}\cdot\textbf{p}\right)^2 + p^4 + p^2 \textbf{k}\cdot\textbf{p})\Big)\bigg)+\mathcal{O}\left(k/m\right)~,
  \end{eqnarray}
\vspace{-1.2cm}
  \begin{eqnarray}
    \mathcal{M}^{\left( \text{s,v} \right)}_{h,ab}(\textbf{k},\textbf{p};t,m) &=& \frac{i}{p^2} \left(-8 k_j \left( \frac{m}{k} \right) \Lambda_{ab}{}^{jc} p_{c} + \left( \frac{k}{m} \right) \frac{4  \Lambda_{ab}{}^{jc} (p^2 - \textbf{k}\cdot\textbf{p}) p_{c}}{k} \right)  +\mathcal{O}\left((k/m)^2\right) ~,\\
    \mathcal{M}^{\left( \text{s,v} \right)}_{V,d}(\textbf{k},\textbf{p};t,m)&=&  8 \left( \frac{m}{k} \right)^2 \frac{1}{p^2} \left(\delta _{d}{}^{j} \textbf{k}\cdot\textbf{p} + k^{j} \left(- \frac{2 k_{d} \textbf{k}\cdot\textbf{p}}{k^2} + p_{d}\right)\right) +\mathcal{O}\left((k/m)^0\right) ~, \\
    \mathcal{M}^{\left( \text{s,v} \right)}_{\psi}(\textbf{k},\textbf{p};t,m)&=&  \frac{i}{p^2} \left(\left( \frac{m}{k} \right) \frac{  k^{j} (k^2 -2 \textbf{k}\cdot\textbf{p})}{k} - \left( \frac{k}{m} \right)\frac{  k^{j} (k^2 -2 \textbf{k}\cdot\textbf{p}) (k^2 + 2 p^2 -2 \textbf{k}\cdot\textbf{p})}{4 k^3}\right)\nonumber \\ && +\mathcal{O}\left((k/m)^2\right) ~, \\
    \mathcal{M}^{\left( \text{s,v} \right)}_{\phi}(\textbf{k},\textbf{p};t,m) &=& \frac{i}{p^2} \Bigg(- \left( \frac{m}{k} \right)^3 \frac{8  k^{j} (k^2 -3 \textbf{k}\cdot\textbf{p})}{k} \nonumber \\ && - \left( \frac{m}{k} \right) \frac{  k^{j} (k^4 + 4 k^2 (p^2 -2 \textbf{k}\cdot\textbf{p}) + 12 \textbf{k}\cdot\textbf{p} (- p^2 + \textbf{k}\cdot\textbf{p}))}{k^3}  \nonumber \\ && + \frac{i}{4} \left( \frac{k}{m} \right) \left( \frac{1}{k} \right)^5 k^{j} (k^6 + 2 k^4 (p^2 -5 \textbf{k}\cdot\textbf{p}) + 4 k^2 (7 \left(\textbf{k}\cdot\textbf{p}\right)^2 + p^4 -4 p^2 \textbf{k}\cdot\textbf{p})  \nonumber \\ && -12 \textbf{k}\cdot\textbf{p} (2 \left(\textbf{k}\cdot\textbf{p}\right)^2 + p^4 -2 p^2 \textbf{k}\cdot\textbf{p}))\Bigg)+\mathcal{O}\left((k/m)^2\right) ~.
  \end{eqnarray}\label{solM}
\end{subequations}
And the function $\textrm{trig}^{(I,J)}_{\delta g}(x)$ is given by 
\begin{subequations}
  \begin{eqnarray}
    \textrm{trig}^{(\text{v,v})}_h(x)&=&\textrm{trig}^{(\text{v,v})}_\psi(x)=\textrm{trig}^{(\text{v,v})}_\phi(x)=\textrm{trig}^{(\text{s,s})}_h(x)\nonumber\\ &=&\textrm{trig}^{(\text{s,s})}_\psi(x)=\textrm{trig}^{(\text{s,s})}_\phi(x)=\textrm{trig}^{(\text{s,v})}_V(x)=\cos(x)~,\\
    \textrm{trig}^{(\text{v,v})}_V(x)&=&\textrm{trig}^{(\text{s,s})}_V(x)=\textrm{trig}^{(\text{v,s})}_h(x)=\textrm{trig}^{(\text{v,s})}_\psi(x)=\textrm{trig}^{(\text{v,s})}_\phi(x)=\sin(x)~.
  \end{eqnarray} \label{trig}
\end{subequations}

\section{Distinctions from pioneering works \label{C}}

The seminal works for deriving the response to vector dark matter in PTAs suggested that the curvature perturbation $\psi$ and the tensor perturbation $h_{ij}$  dominate the PTA signals in the limit $k/m \rightarrow 0$. In this part, we will show how this is incorporated into our framework.

In the limit $k/m \rightarrow 0$, the solution for the vector field in Eqs.~(\ref{8}) and (\ref{solA}) reduces to 
\begin{eqnarray}
  A^{(I)}= \cos(m t)\int \frac{\textrm{d}^3k}{(2\pi)^3}\left\{A_\textbf{k}^{I}e^{-i \textbf{k}\cdot\textbf{x}}\right\} =: \mathcal{A}(\textbf{x})\cos(m t)~. \label{C1}
\end{eqnarray}
Substituting the solution from Eq.~(\ref{C1}) into Eq.~(\ref{EMT}), one can obtain the energy-momentum tensor in the form of
\begin{eqnarray}
T_{bc}^{(v,v)} &=& m^2 A_b^{(v)} A_c^{(v)} - (A_b^{(v)})' (A_c^{(v)})' - \frac{1}{2} \delta_{bc} (m^2 A_j^{(v)} A^{(v),j} - (A_j^{(v)})' (A^{(v),j})')  \nonumber \\
 &=& \frac{1}{2} m^2 \cos(2mt) \left( 2 A_b^{(v)} A_c^{(v)} - \delta_{bc} |\mathcal{A}^{(v)}|^2 \right)
\end{eqnarray}
This result is consistent with Eqs.~(12)--(14) in Ref.~\cite{Nomura:2019cvc}.
 In order to apply the Einstein field equations at the second order $G_{ij}^{(2)} = \kappa T_{bc}^{(v,v)}$, the Einstein tensor can be evaluated as follows,
\begin{subequations}
  \begin{eqnarray}
\delta^{ij} G_{ij}^{(2)} &=& 3\psi'' + \Delta(\phi - \psi) ~,\\
G_{ij}^{(2)} - \frac{1}{3} \delta_{ij} \delta^{ab} G_{ab}^{(2)} &=& \frac{1}{4} (h_{ij}'' - \Delta h_{ij}) + \frac{1}{6} \Delta(\phi - \psi) - \frac{1}{4} (\partial_i V_j' + \partial_j V_i') - \frac{1}{4} \partial_i \partial_j (\phi - \psi)~.
\end{eqnarray}
\end{subequations}
In previous studies \cite{Nomura:2019cvc,Omiya:2023bio}, the authors evaluate the above Einstein tensors to be $\delta^{ij} G_{ij}^{(2)} = 3\psi''+\mathcal{O}(k/m)$ and $G_{ij}^{(2)} - ({1}/{3}) \delta_{ij} \delta^{ab} G_{ab}^{(2)} = ({1}/{4}) h_{ij}''+\mathcal{O}(k/m)$, based on the assumption that spatial derivative terms are all subleading in $k/m$. Thus, only $\psi$ and $h_{ij}$ remain in the equations of motion. Furthermore, based on the solutions of the equations of motion, the redshift from the metric perturbations takes the form of
\begin{eqnarray}
  z = \frac{1}{2}\left(\psi(\textbf{x}_O)-\psi(\textbf{x}_E)\right) +\frac{1}{4}\hat{n}^i \hat{n}^j h_{i j}  \Big|^O_E +\mathcal{O}(k/m)~.
\end{eqnarray}
The timing residual is obtained by integrating the redshift over time $R=\int^t_0 z\textrm{d}t $. Previous works \cite{Nomura:2019cvc,Omiya:2023bio} established that the leading-order PTA response is dominated by the metric perturbations $\psi$ and $h_{ij}$.

In contrast, based on a careful calculation in Sec.~\ref{II}, our results reveal that spatial derivative terms are comparable in magnitude to the time derivatives $\phi''$ and $h_{ij}''$, namely,
\begin{eqnarray}
  \psi'' \simeq h_{ij}'' \simeq \Delta \phi \simeq \partial_i V_j' \simeq \mathcal{O}(m^2)~.\label{C5}
\end{eqnarray}
The explicit expressions for these metric perturbations are provided in Eqs.~(\ref{15a})--(\ref{15d}). This order-of-magnitude estimation based on Eq.~(\ref{C5}) reveals that $\phi$, $h_{ij}\simeq \mathcal{O}(1)$, $\phi \simeq \mathcal{O}\left( (m/k)^2 \right)$ and $V_i \simeq \mathcal{O}(m/k)$. Consequently, the PTA response are dominated by the $\phi$, as shown in Eq.~(\ref{20}), rather than by $\psi$ and $h_{ij}$ as previous claimed. Therefore, our results do not converge to those of Refs.~\cite{Nomura:2019cvc,Omiya:2023bio} under the limit $k/m\rightarrow 0$. This is the key distinction between our study and Refs.~\cite{Nomura:2019cvc,Omiya:2023bio}.

Additionally, we consider the complete  components of the vector field, $A^\mu=(A^{(\text{s})},A^{(\text{v})}_j+\partial_j B^{(\text{s})})$ with $\partial_0 A^{(\text{s})} = \Delta B^{(\text{s})}$, compared to the simplified ansatz $A^\mu=(0,0,0,A_z)$ used in previous studies \cite{Nomura:2019cvc,Omiya:2023bio}. This distinction does not affect our main results.

\section{Power counting in $k/m\rightarrow0$ for metric perturbations and timing residuals }

In addition to the explicit solutions of the metric perturbations in Eq.~(\ref{delta_g}), we can perform power counting with respect to small $k/m$ $\{h_{ab}, V_a, \psi, \phi\}$ by analyzing equations of motion [Eq.~(\ref{Einstein})]. Specifically, in the limit $k/m\rightarrow0$, these metric perturbations  are found to be of the same order of magnitude as 
\begin{subequations}
  \begin{eqnarray}
  \partial_0^2 h_{a b} \sim  \kappa T^{(2)}_{a b} ~, & &
  \partial_0 V_a  \sim  \kappa  \Delta^{-1}\partial^b T^{(2)}_{a b}~, \\ 
  \partial_0^2 \psi \sim  \kappa  T^{(2)}_{a b}~, & &
  \phi  \sim \kappa   \Delta^{-1}\delta^{b c} T^{(2)}_{bc}~.
\end{eqnarray} \label{17}
\end{subequations}
These results are derived under the condition $m\gg k$, which implies $\partial_0 \gg \partial_a$.

Similarly, in the regime $k/m\rightarrow0$, the energy-momentum tensor of the vector field in Eqs.~(\ref{11}) is of the same order of magnitude as
\begin{subequations}
  \begin{eqnarray}
  T_{bc}^{\rm (v,v)} &\sim& m^2 A^{(\rm v)}_b A^{(\rm v)}_c~,\\
  T_{bc}^{\rm (s,s)} &\sim& m^2 \partial_b B^{(\rm s)} \partial_c B^{(\rm s)}= m^2 \Delta^{-1}\partial_b\partial_0 A^{(\rm s)} \Delta^{-1}\partial_c\partial_0 A^{(\rm s)},\\
  T_{bc}^{\rm (s,v)} &\sim& m^2 A_b^{(\rm v)}\partial_c B^{(\rm s)} = m^2 A_b^{(\rm v)} \Delta^{-1}\partial_c \partial_0 A^{(\rm s)}
\end{eqnarray}\label{18}
\end{subequations}
The second equality is a direct consequence of imposing the Lorenz gauge [Eq.~(\ref{LG})].
The Fourier modes of the metric perturbation can be approximately obtained via the substitution  $\partial_0 \rightarrow -i m$, $\partial_j \rightarrow i k_j$ and $\Delta \rightarrow -k^2$. In this context, the Eqs.~(\ref{17}) reduce to $-m^2 h_{ab,\textbf{k}}\sim \kappa T_{ab}^{(2)}$, $ -i m V_{a,\textbf{k}}\sim k^{-2}k^b \kappa T_{ab}^{(2)}$, $-m^2\psi_\textbf{k}\sim \kappa T_{ab}^{(2)}$ and $\phi_\textbf{k}\sim k^{-2}\delta^{ab}\kappa T_{ab}^{(2)}$. 
Similarly, the energy-momentum tensors in Eqs.~(\ref{18}) reduce to $T^{\rm(s,s)}_{bc}\sim -m^4 k^{-4}k_b k_c A^{\rm(s)}A^{\rm(s)}$ and $T^{\rm(s,v)}_{bc}\sim -i m^3 k^{-2}k_c A^{\rm(v)}_b A^{\rm(s)}$.  

Based on the above order-of-magnitude estimates, we summarize the power counting for the metric perturbations in the limit $k/m \rightarrow0$ in Tab.~\ref{T1}.
\begin{table}[ht]
  \centering
  \caption{Order of magnitude estimation for the leading-order metric perturbations $\{h_{ab}, V_a, \psi, \phi\}$. }\label{T1}
\begin{ruledtabular}
  \begin{tabular}{cccccc}
      $\delta g_{\mu\nu}$&&  $T^{(2)}_{b c}$ & $T^{\rm (v,v)}_{b c}/\kappa A^{\rm (v)}_b A^{\rm (v)}_c$ & $T^{\rm (s,s)}_{b c}/\kappa A^{\rm (s)} A^{\rm (s)}$ & $T^{\rm (s,v)}_{b c}/\kappa A^{\rm (s)} A^{\rm (v)}_b$\\ 
    \hline
     $h_{bc,\textbf{k}}$  &$\sim$& $m^{-2} \kappa T^{(2)}_{b c}$ & $(m/k)^0$ &  $(m/k)^2 $  & $(m/k)^1 $ \\
      $V_{b,\textbf{k}}$  &$\sim$& $ (k m)^{-1} \kappa T^{(2)}_{b c}$ & $ (m/k)^1  $ & $(m/k)^3 $ & $(m/k)^2 $\\
      $\psi_\textbf{k}$  &$\sim$& $m^{-2} \kappa T^{(2)}_{b c}$ & $(m/k)^0$& $(m/k)^2 $ & $(m/k)^1 $ \\
      $\phi_\textbf{k}$  &$\sim$ & $k^{-2} \kappa T^{(2)}_{b c}$ & $(m/k)^2 $ & $(m/k)^4 $ & $(m/k)^3$ \\
\end{tabular}
\end{ruledtabular}
\end{table}
It shows that the power counting in $k/m$ for the metric perturbations is consistent with the leading-order results in Eqs.~(\ref{delta_g}). 

Based on Eqs.~(\ref{zgeneral})--(\ref{22eq}) and Eqs.~(\ref{17})--(\ref{18}), we can present the power counting for the leading-order PTA observable $z$ via
\begin{subequations}
  \begin{eqnarray}
  z_{h} \sim \hat{n}^a\hat{n}^b h_{ab}~, & & z_V \sim \frac{\textbf{k}\cdot\hat{n}}{2\pi f+\textbf{k}\cdot\hat{n}}\hat{n}^a V_a~, \\
  z_\psi \sim \psi~, & & z_\phi \sim \frac{\textbf{k}\cdot\hat{n}}{2\pi f+\textbf{k}\cdot\hat{n}} \phi~. 
\end{eqnarray} 
\end{subequations}
Associating the results in Tab.~\ref{T1} and  ${\textbf{k}\cdot\hat{n}}/{(2\pi f+\textbf{k}\cdot\hat{n})}\simeq k/m$, we can obtain $(k/m)z_\phi \sim z_\psi \sim z_h \sim z_V$.
Namely, the PTA observable $z$ in Eq.~(\ref{z}) is dominated by $\phi$ in the limit of $k/m \rightarrow0$. We summarize the order-of-magnitude estimates for the $z$ in Tab.~\ref{T2}.
\begin{table}[ht]
  \centering
  \caption{Order of magnitude estimation for the leading-order PTA observable $z$. }\label{T2}
\begin{ruledtabular}
  \begin{tabular}{cccc}
      & (v,v) & (s,s) &(s,v)\\ 
    \hline
     $z_h$  & $(m/k)^0$ &  $(m/k)^2 $  & $(m/k)^1 $ \\
      $z_V$  & $ (m/k)^0  $ & $(m/k)^2 $ & $(m/k)^1 $\\
      $z_\psi$  & $(m/k)^0$& $(m/k)^2 $ & $(m/k)^1 $ \\
      $z_\phi$  & $(m/k)^1 $ & $(m/k)^3 $ & $(m/k)^2$ \\
\end{tabular}
\end{ruledtabular}
\end{table}

\section{Density of vector dark matter \label{appB}}
The density of vector dark matter can be given with the time-time component of energy-momentum tensor. Specifically, we have $\rho=T_{00}$, and
\begin{eqnarray}
  \rho &=& \rho^{(\text{s,s})} +\rho^{(\text{s,v})}+\rho^{(\text{v,v})} \nonumber \\
       &=& \frac{1}{2}\partial_j A^{(\text{s})} \partial^j A^{(\text{s})}-\partial_j \partial_0 B^{(\text{s})} \partial^j A^{(\text{s})}+ \frac{1}{2}\partial_j \partial_0 B^{(\text{s})} \partial^j \partial_0 B^{(\text{s})} + \frac{1}{2}m^2\left( \left(A^{(\text{s})}  \right)^2 + \partial_j B^{(\text{s})} \partial^j B^{(\text{s})}\right) \nonumber\\ 
       & & +m^2 A^{(\text{v}),j}\partial_j B^{(\text{s})} +  \partial_0 A^{(\text{s}),j}\left( \partial_j\partial_0 B^{(\text{s})} -\partial_j A^{(\text{s})} \right) \nonumber \\ 
       & & +\frac{1}{2} \partial_0 A_j^{(\text{v})} \partial_0 A^{(\text{v}),j}-\frac{1}{2} \partial_j A^{(\text{v})}_l \partial^l A^{(\text{v}),j}+\frac{1}{2}\partial_j A^{(\text{v})}_l \partial^j A^{(\text{v}),l}+\frac{1}{2}m^2 A^{(\text{v}),j} A^{(\text{v})}_j~.
\end{eqnarray}
By making use of Lorentz gauge in Eq.~(\ref{LG}) and solution of vector field in Eq.~(\ref{solA}), the Fourier modes of the  densities are evaluated to be
\begin{subequations}
  \begin{eqnarray}
  \rho_\textbf{k}^{\text{(s,s)}} &=& m^2 \int \frac{\textrm{d}^3p}{(2\pi)^3}\Bigg\{\frac{\tilde{A}^{\text{(s)}}_{\textbf{k}-\textbf{p}}\tilde{A}_\textbf{p}^{\text{(s)}}}{4p^2|\textbf{k}-\textbf{p}|^2}\Big( 3k^2p^2-k\cdot p(k^2+4p^2)+2(k\cdot p)^2 \nonumber \\ && +(p^2 - k \cdot p)(w_\textbf{p} -w_{\textbf{k}-\textbf{p}})^2 \Big)  \Bigg\}~, \\
  \rho_\textbf{k}^{\text{(s,v)}} &=&   \int \frac{\textrm{d}^3p}{(2\pi)^3}\left\{\tilde{A}^{\text{(s)}}_{\textbf{k}-\textbf{p}}\tilde{A}_\textbf{p}^{\text{(v)},b} p_b \left(\frac{m}{p}\right)^2 (w_{\textbf{k}-\textbf{p}} - w_{\textbf{p}})\right\}~,\\
  \rho_\textbf{k}^{\text{(v,v)}} &=& \frac{1}{4} \int \frac{\textrm{d}^3p}{(2\pi)^3}\Bigg\{\tilde{A}^{\text{(v)},a}_{\textbf{k}-\textbf{p}}\tilde{A}_\textbf{p}^{\text{(v)},b} \Big( ((w_{\textbf{k}-\textbf{p}} - w_{\textbf{p}})^2-k^2)\delta_{ab}+2p_a k_b  \Big) \Bigg\}~.
\end{eqnarray}  
\end{subequations}
For the deterministic sources given in Eqs.~(\ref{Dsrc}), the density reduce to
\begin{subequations}
    \begin{eqnarray}
  \rho_{2\textbf{k}_\ast}^{\text{(s,s)}} &=& 2(2\pi)^3m^2\left(\mathcal{A}_{\textbf{k}_\ast}^{\text{(s)}}\right)^2(1-\cos(2w_{k_\ast} t))\delta^3(\textbf{k}-2\textbf{k}_\ast)  ~,  \\
  \rho_{2\textbf{k}_\ast}^{\text{(s,v)}} &=&   0~,  \\
  \rho_{2\textbf{k}_\ast}^{\text{(v,v)}} &=& -2(2\pi)^3 k^2_\ast\left|\mathcal{A}_{\textbf{k}_\ast}^{\text{(v)}}\right|^2 (1-\cos(2w_{k_\ast} t)) \delta^3(\textbf{k}-2\textbf{k}_\ast) 
\end{eqnarray}\label{B3}
\end{subequations}
Here, the density of vector dark matter at wavenumber $\textbf{k}_\ast$ do not explicitly diverge as $k_\ast/m \rightarrow0$. This contrasts with the timing residuals in Eq.~(\ref{R}), where the leading-order amplitude is proportional to negative powers of $k_\ast/m$. It indicates that our calculation is physically valid: the small dark speed ($v\approx10^{-3}$) improve the observational feasibility in PTA observations, and does not result in enhancement of the dark matter density in our galaxy. Additionally, we wish to point out that the metric perturbations proportional to negative power of $k/m$ were neglected in Refs.~\cite{Nomura:2019cvc,Omiya:2023bio}. In these studies, leading-order timing residual amplitude is thus proportional to $\mathcal{O}\left(\left(k/m\right)^0\right)$, and does not exhibit the aforementioned features. We bring the missing terms back through a rigorous derivation with a finite $k/m$, and show their dominant contributions to the timing residuals. 

Since the density oscillates with time, it inevitably affects the timing residuals. From time-time component of Einstein field equations, we have $\Delta \psi = \kappa \rho$. Associating this equation with Eqs.~(\ref{B3}), we can obtain $\psi^{\text{(s,s)}}_\textbf{k}\propto \left((m/k)^2\right)$ and $\psi^{\text{(v,v)}}_\textbf{k}\propto \mathcal{O}(1)$ at the leading order, which are consistent with our results in Eqs.~(\ref{15c}) and (\ref{15g}).

\bibliography{cite}
\end{document}